\documentclass[twocolumn]{aastex63}

\usepackage{indentfirst}
\usepackage{appendix}
\usepackage{makecell}
\usepackage{tabu}
\usepackage{float}
\usepackage{lineno}

\DeclareUnicodeCharacter{2212}{-}
\DeclareUnicodeCharacter{00B4}{'}
\DeclareUnicodeCharacter{00A8}{"}

\shorttitle{Spectroscopic confirmation of dual quasars}
\shortauthors{Tang et al.}

\graphicspath{{./}}

\begin{document}

\title{Optical Spectroscopy of Dual Quasar Candidates from the Subaru HSC-SSP program}

\author[0000-0002-2185-5679]{Shenli Tang}
\affiliation{Department of Physics, University of Tokyo, Tokyo 113-0033, Japan}
\affiliation{Kavli Institute for the Physics and Mathematics of the Universe (WPI), The University of Tokyo, Kashiwa, Chiba 277-8583, Japan}
\affiliation{Institute for Cosmic Ray Research, The University of Tokyo, 5-1-5 Kashiwanoha, Kashiwa, Chiba 277-8582, Japan}

\author[0000-0002-0000-6977]{John D. Silverman}
\affiliation{Kavli Institute for the Physics and Mathematics of the Universe (WPI), The University of Tokyo, Kashiwa, Chiba 277-8583, Japan}
\affiliation{Department of Astronomy, School of Science, The University of Tokyo, 7-3-1 Hongo, Bunkyo, Tokyo 113-0033, Japan}

\author[0000-0001-8917-2148]{Xuheng Ding}
\affiliation{Kavli Institute for the Physics and Mathematics of the Universe (WPI), The University of Tokyo, Kashiwa, Chiba 277-8583, Japan}

\author{Junyao Li}
\affiliation{CAS Key Laboratory for Research in Galaxies and Cosmology, Department of Astronomy, University of Science and Technology of China, Hefei 230026, China}
\affiliation{School of Astronomy and Space Science, University of Science and Technology of China, Hefei 230026, China}
\affiliation{Kavli Institute for the Physics and Mathematics of the Universe (WPI), The University of Tokyo, Kashiwa, Chiba 277-8583, Japan}

\author[0000-0001-9299-5719]{Khee-Gan Lee}
\affiliation{Kavli Institute for the Physics and Mathematics of the Universe (WPI), The University of Tokyo, Kashiwa, Chiba 277-8583, Japan}

\author[0000-0002-0106-7755]{Michael A. Strauss}
\affiliation{Department of Astrophysical Sciences, Princeton University, 4 Ivy Lane, Princeton, NJ 08544, USA}

\author[0000-0003-4700-663X]{Andy Goulding}
\affiliation{Department of Astrophysical Sciences, Princeton University, 4 Ivy Lane, Princeton, NJ 08544, USA}

\author{Malte Schramm}
\affiliation{National Astronomical Observatory of Japan, 2-21-1 Osawa, Mitaka, Tokyo 181-8588, Japan}

\author{Lalitwadee Kawinwanichakij}
\affiliation{Kavli Institute for the Physics and Mathematics of the Universe (WPI), The University of Tokyo, Kashiwa, Chiba 277-8583, Japan}

\author[0000-0002-7738-6875]{J. Xavier Prochaska}
\affiliation{University of California Observatories–Lick Observatory, University of California, Santa Cruz, CA 95064, USA}
\affiliation{Kavli Institute for the Physics and Mathematics of the Universe (WPI), The University of Tokyo, Kashiwa, Chiba 277-8583, Japan}

\author[0000-0002-7054-4332]{Joseph F. Hennawi}
\affiliation{University of California Santa Barbara, Santa Barbara, CA, USA}

\author[0000-0001-6186-8792]{Masatoshi Imanishi}
\affiliation{National Astronomical Observatory of Japan, 2-21-1 Osawa, Mitaka, Tokyo 181-8588, Japan}

\author[0000-0002-4923-3281]{Kazushi Iwasawa}
\affiliation{ICREA and Institut de Ciències del Cosmos, Universitat de Barcelona, IEEC-UB, Martí i Franquès, 1,08028 Barcelona, Spain}

\author[0000-0002-3531-7863]{Yoshiki Toba}
\affiliation{Department of Astronomy, Kyoto University, Kitashirakawa-Oiwake-cho, Sakyo-ku, Kyoto 606-8502, Japan}
\affiliation{Academia Sinica Institute of Astronomy and Astrophysics, 11F of Astronomy-Mathematics Building, AS/NTU, No.1, Section 4, Roosevelt Road, Taipei 10617, Taiwan}
\affiliation{Research Center for Space and Cosmic Evolution, Ehime University, 2-5 Bunkyo-cho, Matsuyama, Ehime 790-8577, Japan}

\author{Issha Kayo}
\affiliation{Department of Liberal Arts, Tokyo University of Technology, Ota-ku, Tokyo 144-8650, Japan}

\author[0000-0003-3484-399X]{Masamune Oguri}
\affiliation{Department of Physics, University of Tokyo, Tokyo 113-0033, Japan}
\affiliation{Kavli Institute for the Physics and Mathematics of the Universe (WPI), The University of Tokyo, Kashiwa, Chiba 277-8583, Japan}
\affiliation{Research Center for the Early Universe, University of Tokyo, Tokyo 113-0033, Japan}

\author{Yoshiki Matsuoka}
\affiliation{Research Center for Space and Cosmic Evolution, Ehime University, 2-5 Bunkyo-cho, Matsuyama, Ehime 790-8577, Japan}

\author[0000-0002-4377-903X]{Kohei Ichikawa}
\affiliation{Frontier Research Institute for Interdisciplinary Sciences, Tohoku University, Sendai 980-8578, Japan}

\author[0000-0001-6742-8843]{Tilman Hartwig}
\affiliation{Institute for Physics of Intelligence, School of Science, The University of Tokyo, Bunkyo, Tokyo 113-0033, Japan}

\author[0000-0001-5493-6259]{Nobunari Kashikawa}
\affiliation{Department of Astronomy, School of Science, The University of Tokyo, 7-3-1 Hongo, Bunkyo, Tokyo 113-0033, Japan}

\author{Toshihiro Kawaguchi}
\affiliation{Department of Economics, Management and Information Science, Onomichi City University, Hisayamada 1600-2, Onomichi, Hiroshima 722- 8506, Japan}

\author[0000-0002-4052-2394]{Kotaro Kohno}
\affiliation{Department of Astronomy, School of Science, The University of Tokyo, 7-3-1 Hongo, Bunkyo, Tokyo 113-0033, Japan}

\author[0000-0003-1747-2891]{Yuichi Matsuda}
\affiliation{National Astronomical Observatory of Japan, 2-21-1 Osawa, Mitaka, Tokyo 181-8588, Japan}

\author{Tohru Nagao}
\affiliation{Research Center for Space and Cosmic Evolution, Ehime University, 2-5 Bunkyo-cho, Matsuyama, Ehime 790-8577, Japan}

\author[0000-0001-9011-7605]{Yoshiaki Ono}
\affiliation{Institute for Cosmic Ray Research, The University of Tokyo, 5-1-5 Kashiwanoha, Kashiwa, Chiba 277-8582, Japan}

\author[0000-0003-2984-6803]{Masafusa Onoue}
\affiliation{Max-Planck-Institut für Astronomie, Königstuhl 17, D-69117 Heidelberg, Germany}

\author[0000-0002-1049-6658]{Masami Ouchi}
\affiliation{Institute for Cosmic Ray Research, The University of Tokyo, 5-1-5 Kashiwanoha, Kashiwa, Chiba 277-8582, Japan}
\affiliation{National Astronomical Observatory of Japan, 2-21-1 Osawa, Mitaka, Tokyo 181-8588, Japan}

\author[0000-0002-2597-2231]{Kazuhiro Shimasaku}
\affiliation{Department of Astronomy, School of Science, The University of Tokyo, 7-3-1 Hongo, Bunkyo, Tokyo 113-0033, Japan}

\author[0000-0002-2536-1633]{Hyewon Suh}
\affiliation{Subaru Telescope, National Astronomical Observatory of Japan (NAOJ), National Institutes of Natural Sciences (NINS), 650 North A’ohoku place, Hilo, HI 96720, USA}

\author[0000-0001-7266-930X]{Nao Suzuki}
\affiliation{Kavli Institute for the Physics and Mathematics of the Universe (WPI), The University of Tokyo, Kashiwa, Chiba 277-8583, Japan}

\author[0000-0003-2247-3741]{Yoshiaki Taniguchi}
\affiliation{The Open University of Japan, 2-11 Wakaba, Mihama-ku, Chiba 261-8586, Japan}

\author[0000-0001-7821-6715]{Yoshihiro Ueda}
\affiliation{Department of Astronomy, Kyoto University, Kitashirakawa-Oiwake-cho, Sakyo-ku, Kyoto 606-8502, Japan}

\author{Naoki Yasuda}
\affiliation{Kavli Institute for the Physics and Mathematics of the Universe (WPI), The University of Tokyo, Kashiwa, Chiba 277-8583, Japan}

\begin{abstract}


We report on a spectroscopic program to search for dual quasars using Subaru Hyper Suprime-Cam (HSC) images of SDSS quasars which represent an important stage during galaxy mergers. Using Subaru/FOCAS and Gemini-N/GMOS, we identify three new physically associated quasar pairs having projected separations less than 20 kpc, out of 26 observed candidates. These include the discovery of the highest redshift ($z=3.1$) quasar pair with a separation $<$ 10 kpc. Based on the sample acquired to date, the success rate of identifying physically associated dual quasars is $19\%$ when excluding stars based on their HSC colors. Using the full sample of six spectroscopically confirmed dual quasars, we find that the black holes in these systems have black hole masses ($M_{BH} \sim 10^{8-9}M_{\odot}$) similar to single SDSS quasars as well as their bolometric luminosities and Eddington ratios. We measure the stellar mass of their host galaxies based on 2D image decomposition of the five-band ($grizy$) optical emission and assess the mass relation between supermassive black holes (SMBHs) and their hosts. Dual SMBHs appear to have elevated masses relative to their host galaxies. Thus mergers may not necessarily align such systems onto the local mass relation, as suggested by the Horizon-AGN simulation. This study suggests that dual luminous quasars are triggered prior to the final coalescence of the two SMBHs, resulting in early mass growth of the black holes relative to their host galaxies.

\end{abstract}

\keywords{galaxies: active – galaxies: interactions – galaxies: nuclei}

\section{Introduction} 
It is widely accepted that supermassive black holes (SMBHs; mass $> 10^6 M_{\odot}$) grow from an accretion disk, likely replenished with interstellar gas, and release prodigious levels of energy, thus working as an engine to power the class of objects known as quasars and active galactic nuclei (AGNs) \citep{hoyle1962nature, salpeter1964accretion, lynden1969galactic, zel1989fate}. 
\cite{soltan1982masses} argued that if quasars were powered by accretion onto a SMBH, then such SMBHs must exist in our local universe as ``dead'' quasars, thus the local BH density can be estimated via the integrated quasar luminosity. 

Spectroscopy of the nuclei of nearby galaxies with the Hubble Space Telescope (HST) has found evidence for SMBHs in many galaxies, leading to the convincing conclusion that BHs are present not only in AGNs, but also many normal galaxies \citep{kormendy1995inward,kormendy2001supermassive}. This attracted interest on studies of the correlation between the BHs and their host galaxies \citep[reviewed by ][]{kormendy2013coevolution}. It turns out that the mass of the SMBHs correlate with various host properties, such as stellar velocity dispersion ($\sigma_*$) luminosity ($L_{\mathrm{host}}$) and stellar mass ($M_*$) \citep{magorrian1998demography,ferrarese2000fundamental,gebhardt2000black,haring2004black,beifiori2012correlations}. 

As a result, there should be a way for the host galaxy to communicate with the central SMBH or vice versa. A direct way is that the host galaxy transfers its matter to the SMBH via some mechanism. But it is unclear how 
matter could lose angular momentum from galaxy-wide scales to fall into the central BH. Numerical simulations have shown that interactions and mergers between galaxies could trigger large-scale gas inflows. Such processes will feed the central SMBH and power it as a quasar \citep{di2005energy}. Multi-wavelength observations have provided evidence that the interaction of galaxies do, to some level, ignite AGNs \citep{silverman2011impact, ellison2011galaxy, mechtley2016most, goulding2018galaxy, koss2018population}. While such interactions mainly trigger high luminosity AGNs, the low luminosity AGNs are thought to be triggered by secular processes such as explored by \cite{ciotti1997cooling,ciotti2001cooling}.
\par
For some time, there has been much interest in confirming the merger scenario for fueling SMBHs by studying the occurrence of simultaneous accretion events in each constituent galaxy in a merger. \cite{begelman1980massive} reveals the possibility for such SMBH binaries to exist, originating from two interacting gas-rich galaxies in which dynamical friction of the gas brings the cores of the galaxies down to pc scale separation with a characteristic time $\sim10^8$ years \citep{yu2002observational}. Further hardening of the black hole binary is through stellar diffusion and gravitational waves. More recently, numerical simulations (\cite{hopkins2006unified} \cite{capelo2015growth} and \cite{capelo2017survey}) have shown that in a merger phase, both SMBHs could be activated well before final coalescence. While quasar activity can be highly obscured, they become subsequently visible after feedback from the central SMBH expels gas and dust.
\par
In recent decades, a number of systematic studies have been carried out to search for binary quasars, including some serendipitous findings \citep[e.g. ][]{bianchi2008chandra,green2010sdss,huang2014hst}. \cite{de2020quest} provides a panoramic review on these efforts to search for binary quasars. Such studies have shed light on quasar clustering and quasar-galaxy clustering on smaller scales. For example, \cite{hennawi2006binary} discovered 221 quasar pairs over the redshift range $0.5 \sim 3.0$ from the SDSS and the 2dF quasar Redshift Survey via color selection and spectroscopic data, and calculated the quasar correlation function from $\sim 400h^{-1}$ kpc down to $\sim10\,h^{-1}$ kpc.  \cite{hennawi2010binary} and \cite{shen2010binary} expanded the sample to $z > 2.9$, finding strong small-scale clustering at these high redshifts.  Close quasar pairs are often interpreted to be sites of galaxy interactions \citep{djorgovski1991quasars}. Recently, \cite{eftekharzadeh2017clustering} reported a steeper power-law index of $\gamma = 1.97 \pm 0.03$ for the correlation function while fixing $r_0 = 5\,h^{-1}$ cMpc, compared to 1.6 $\sim$ 2.0 for galaxies in the literature \citep[e.g. ][]{coil2017primus,masjedi2006very,zehavi2011galaxy,zhai2016clustering}, which again, supported the positive influence of local environment (i.e., mergers) on black hole growth. 
However, these studies have been based on the SDSS, which is limited by the seeing of the SDSS imaging (typically 1.1-1.4 arcsec) and the aperture of the SDSS spectroscopic fibers (3\arcsec\  for the first phases of SDSS; 2\arcsec\ thereafter), limiting the separations they could probe to these levels. 
\par
Previous searches have tried to break the constraint on spatial resolution via a further analysis of the spectroscopic data. For example, \cite{liu2010discovery} suggested that dual AGN
may possess double-peaked [O\,{\sc iii}]$\lambda\lambda$4959,5007 lines, and successfully confirmed four kpc-scale AGN pairs originally selected from SDSS. However, follow-up studies with near-infrared (NIR) imaging and optical slit spectroscopy \citep{shen2011type} revealed that only a small fraction ($\sim10\%$) of such objects are best explained by binary quasars, with most of them formed by kinematics of single AGNs.  \cite{comerford2013dual} carried out a similar study by selecting AGNs with double-peaked narrow emission lines from the AGN and Galaxy Evolution Survey (AGES). Follow-up with X-ray observations was only able to confirm one object as a binary quasar \citep{comerford2015merger}.
\par
Even so, SMBH binaries (SMBHBs) are expected to be the brightest sources of nanohertz gravitational waves (GWs) in their final stage of evolution, which should be detectable by pulsar timing arrays \citep[PTAs, ][]{burke2019astrophysics,arzoumanian2021nanograv}. One needs an estimate of the space density of close massive SMBH pairs to estimate how strong the GW background is
\citep{inayoshi2018gravitational}. \cite{goulding2019discovery} discovered a dual quasar at $z\sim0.198$ separated by only $\sim$430 pc. The target was selected from a study of AGN having narrow emission lines and characterized as having strong asymmetries in its [O\,{\sc iii}] $\lambda$5007 emission \citep{mullaney2013narrow}. They found that the [O\,{\sc iii}] and continuum emission was coming from two distinct sources using data from both HST and Chandra, making it the most luminous dual quasar known in the relatively nearby universe. 
\par
We have begun a program using Subaru HSC imaging to search for dual quasars at close separation (down to a few kpc), and study their physical properties including their BH masses, host masses, and Eddington ratios. In \citet{silverman2020dual}, we reported three spectroscopically confirmed quasar pairs using Keck/LRIS and discussed the sample selection and methodology. 
Based on the number of dual quasar candidates and the fraction of objects confirmed as dual quasars from the spectroscopy, we reported a dual fraction of 0.26$\%$, with no evidence for evolution with redshift. However, this was based solely on a success rate determined at $z<1$. Our aim in this paper is to build a larger sample and identify the first spectroscopically confirmed dual quasars from our sample at $z>1$.

\par
\begin{figure}[htb]
\hspace{-2em}
\includegraphics[width=90mm]{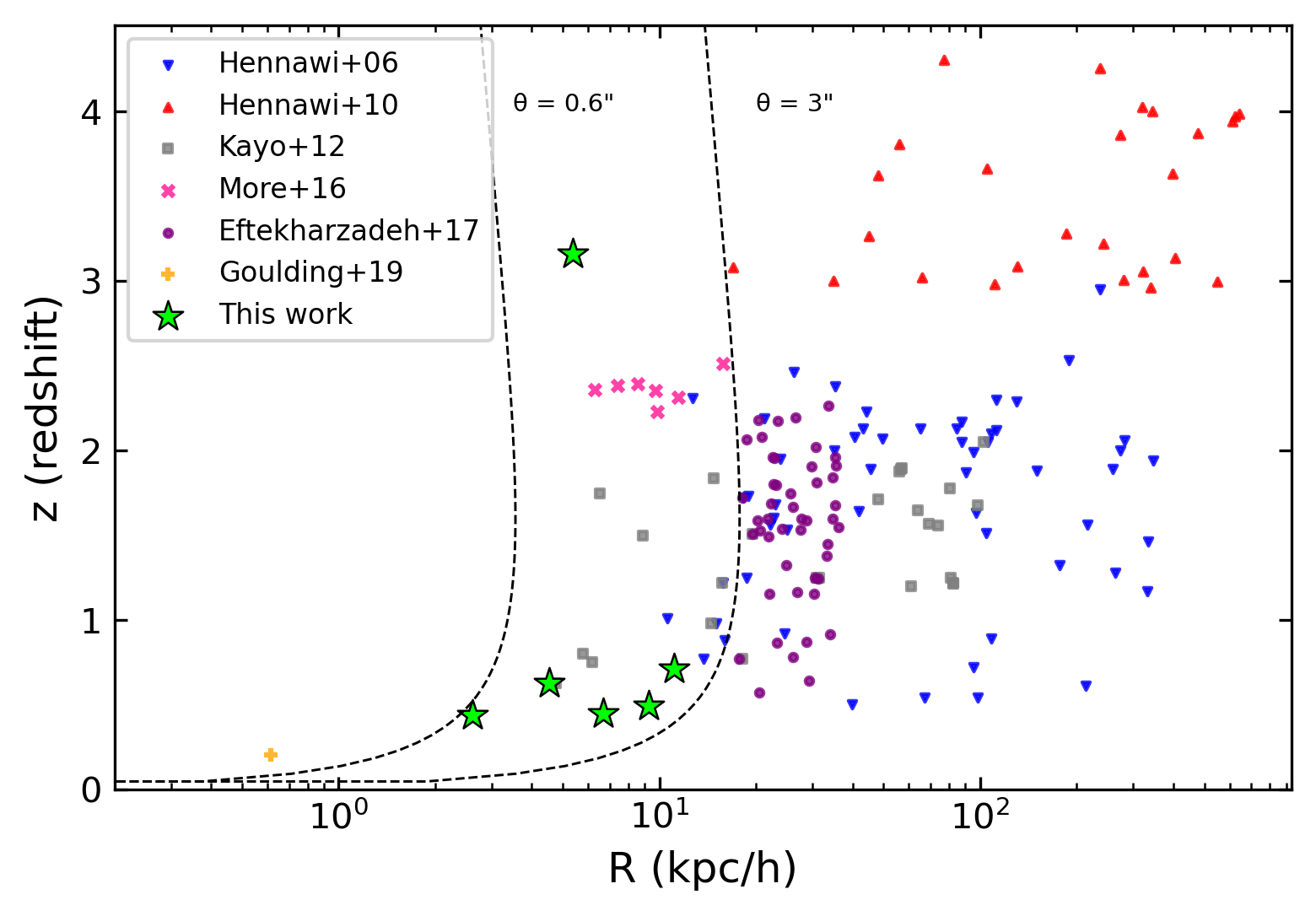}
\caption{Distribution of dual quasars in the redshift--projected separation space (in kpc). The dashed lines indicate the region of our selected candidates with the angular separation given in arc-seconds. Green stars mark are newly discovered quasar pairs in our study. Comparison dual quasars are from \cite{hennawi2006binary}, \cite{hennawi2010binary}, \cite{kayo2012very}, \cite{more2016sdss}, \cite{eftekharzadeh2017clustering} and \cite{goulding2019discovery}.
\label{fig:sep_dis}}
\end{figure}

Figure \ref{fig:sep_dis} shows the distribution of redshift versus projected separation of dual quasars based on an up-to-date compilation of surveys of dual quasars including the objects we confirm in this paper. Compared to previous studies at either larger separation or over a narrow redshift interval, the region between the dashed lines includes very few confirmed dual quasars. This parameter space is our target, i.e., between 0\arcsec.6 and 3\arcsec~extending from redshift 0 to 4.5. In this work, we first give a quick review of our sample selection in \textsection \ref{sec:Finding}. 
The details of spectroscopic follow-up observations using Subaru/FOCAS and Gemini/GMOS-N are described in \textsection \ref{sec:Spec}. We then fit the spectra according to the models described in \textsection \ref{sec:Fitting}. \textsection \ref{sec:Results} summarizes the results of this program, including the success rate and details of individual targets. \textsection \ref{sec:Discussion} presents the physical properties of these objects based on spectral fitting and image decomposition, and compares the results with other studies. \textsection \ref{sec:Conclusion} summarizes the findings of this work and compares our results with a 
matched sample from the Horizon-AGN simulation. Throughout this paper, we adopt a standard $\mathrm{\Lambda CDM}$ cosmology model with $\mathrm{H_0} = 70\,\mathrm{km\,s^{-1}} \mathrm{Mpc^{−1}}$, $\Omega_m = 0.30$, and $\Omega_{\Lambda} = 0.70$, so that an angular separation of 1\arcsec at redshift 1.5 corresponds to a projected distance 5.92 kpc.

\section{Finding Dual Quasar Candidates with HSC} \label{sec:Finding}
\subsection{SDSS DR14}
The Sloan Digital Sky Survey (SDSS) uses a 2.5-m wide-angle optical telescope at Apache Point Observatory (APO) in New Mexico. Since the start of operations in 2000, its spectral database has grown to 4 million objects covering over 14,055 square degrees of sky \citep[up to DR14, ][]{abolfathi2018fourteenth,paris2018sloan}. DR14 is the second data release of the fourth phase of the Sloan Digital Sky Survey (SDSS-IV). It provides a quasar catalog from the extended Baryon Oscillation Spectroscopic Survey (eBOSS), that also includes previously spectroscopically confirmed quasars from SDSS-I, II \citep{schneider2010sloan} and III \citep{eisenstein2011sdss,paris2012sloan,paris2017sloan}. The SDSS quasar selection includes magnitude- and color-selected samples and objects identified through radio, infrared and X-ray surveys including FIRST, GALEX, 2MASS and ROSAT. In total, the catalog contains 526,356 quasars up to $z \sim 5$. The spectral data are taken with fibers, each subtending 3\arcsec\ (SDSS-I/II) or 2\arcsec\ (BOSS/eBOSS) diameter on the sky, connected to plates which each have a field of view of approximately 7 $\mathrm{deg^2}$ and connecting 640 (SDSS-I/II) or 1000 (BOSS/eBOSS)fibers. The earlier spectroscopic data (SDSS-I/II) covers wavelengths from 3800 to 9200 \text{\AA} at a spectral resolution of $\sim$ 2000 \citep{schneider2010sloan}. The later data (BOSS/eBOSS) cover 3600 to 10,400 \text{\AA} with resolution varying from $\sim$ 1300 at 3600 \text{\AA} to $\sim$ 2500 at 10,000 \text{\AA} \citep{paris2018sloan}.
\par
\subsection{Subaru HSC}
Hyper Suprime-Cam (HSC, \cite{miyazaki2018hyper}) is a wide-field (1.7 degree diameter) optical imager installed at the prime focus of the Subaru Telescope. The Subaru Strategic Program (SSP, \cite{aihara2018hyper}) is a survey using this instrument that began in March 2014 with the second public data release in 2019 (PDR2, \cite{aihara2019second}). The survey includes three layers of imaging data: Wide, Deep, and UltraDeep. The imaging data used in this study is from the Wide layer with 796 $\rm{deg^2}$ of $i$-band imaging having at least one 200 second exposure, with median seeing of 0.58 arcsec and depth 26.2 mag for $5\sigma$ detection. Our dual quasar search is first based on the $i$-band. We then include the images taken in the other four broad-band filters (i.e., $g$, $r$, $z$, $y$; \cite{kawanomoto2018hyper}) to provide color information and stellar mass estimation for objects at $z\lesssim1$.
\par
\subsection{Candidate Selection} \label{subsec:Candidates}
Since HSC does not yet have a sample of spectroscopically-confirmed quasars based on optical selection, we chose to start our search for dual quasars using the catalog of spectroscopically confirmed SDSS quasars \citep{paris2018sloan}. This catalog provides additional  information such as emission line properties which can be used to estimate black hole masses. To search for dual quasars, the SDSS DR14 quasar catalog ensures that at least one has already been identified. Our idea is to identify those that have one (or more) close companions, but only initially counted as a single object because of the limitation of the SDSS spatial resolution, fiber collisions, or faintness of the optical companion.
\par
\begin{figure}[htp]
\hspace{-2em}
\includegraphics[width=90mm]{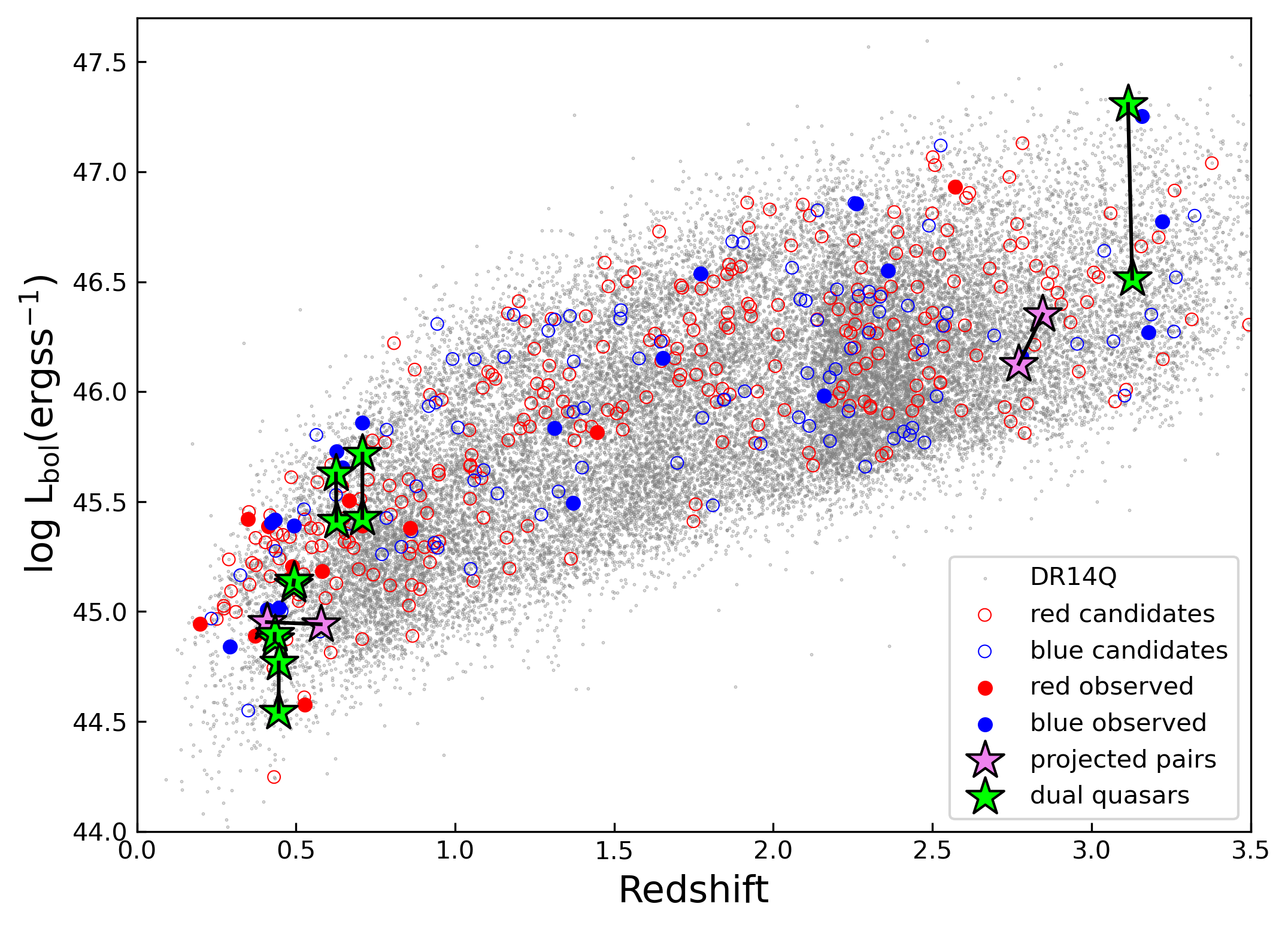}
\caption{Distribution of dual quasar candidates in the bolometric luminosity versus redshift plane. Grey dots are 34,476 SDSS DR14 quasars, imaged by HSC. Colored circles are 401 dual quasar candidates where red open circles have $g - r > 1.0$ and blue open circles have $g - r \leq 1.0$. The filled circles are objects whose spectroscopy reveals than not to be dual quasars. 
Green stars mark the successfully confirmed dual quasars. Pink stars are projected quasar pairs with two sources at different redshifts. We connect the two objects of the confirmed and projected pairs with 
solid black lines.
\label{fig:bol_z}}
\end{figure}

We match the SDSS quasar catalog with the HSC photometric catalog to make use of the higher resolution and deeper optical imaging available through the HSC SSP. There are 34,476 SDSS quasars imaged by HSC that are unsaturated ($i_{\mathrm{AB}} \geq 18$), have no  bad pixels, and have model magnitudes available. 
HSC image cutouts of size $60^{\prime\prime}$ $\times$ $60^{\prime\prime}$ are generated for each object in each band, together with the variance image and model point-spread function (PSF). As described in \cite{silverman2020dual}, we run a Lenstronomy-based \citep{birrer2018lenstronomy} modeling algorithm on the images with a composition of point sources and Sersic profiles. We then pick out candidates that appear to have two or more point sources within the field, with separation ranging from $0^{\prime\prime}.6$ to $4^{\prime\prime}$. After filtering out some unlikely cases and known lenses \citep{inada2008sloan, inada2010sloan, inada2012sloan}, there are 421 candidates left, of which 401 have photometry in all five bands.  We will focus on those companion objects that have colors suggestive of being quasars, to remove stellar contamination. 
Figure \ref{fig:bol_z} shows the distribution of bolometric luminosity and redshift for the 401 dual quasar candidates and the parent population of SDSS quasars having HSC imaging. We separate our candidates into blue ($g-r\leq1$) and red ( $g-r>1$) companions as shown in the figure. We give higher priority to the blue sources in the color-color diagram to remove objects that are likely to be stars
(see Section \ref{sec:Fitting} for details) when designing our observations. In total, we have obtained spatially resolved spectra of 32 objects. 
\par

\section{Spectroscopic acquisition, data reduction, and classification} \label{sec:Spec}
As each object in our sample was drawn from the SDSS DR14 quasar catalog, we know that each pair includes at least one quasar. 
While our decomposition algorithm identifies point sources, the companion still could be a star, a galaxy (the ``point source" may turn out to be the bulge), or a quasar at a different redshift. Therefore, we need to follow up with spectroscopic observations to confirm the nature of the companions and to determine the physical properties of the sources. 

Our spectroscopic observations are carried out 
in long slit mode with Subaru/FOCAS and Gemini/GMOS-N. Most of our spectroscopy separates the two sources in each pair well. Two separate gaussian kernels can be identified from the spatial profiles. While the kernels might overlap in some cases, spatial profile fitting is applied to deal with blending (see Figure \ref{fig:0219} for an example). However, in one source,  232853.40+011221.8, the spectra are highly blended and the [O\,{\sc iii}] $\lambda$5007 profile of the main source is extremely extended, so we are not able to deblend the two objects. The basic properties of the targets are shown in Table \ref{tab:basic}, and the observational setups are shown in Table \ref{tab:observation}.

\begin{deluxetable*}{l|lcccccccccccc}
\tablenum{1}
\tablecaption{Basic information\label{tab:basic}}{}
\tablewidth{0pt}
\tablehead{
\colhead{} & \colhead{Name} & \colhead{RA1} & \colhead{DEC1} & \colhead{RA2} & \colhead{DEC2} & \colhead{Redshift} & \colhead{Sep} & \colhead{mag$_g$1} & \colhead{mag$_r$1} & \colhead{mag$_i$1} & \colhead{mag$_g$2} & \colhead{mag$_r$2} & \colhead{mag$_i$2}\\
\colhead{} & \colhead{(J2000)} & \colhead{(degree)} & \colhead{(degree)} & \colhead{(degree)} & \colhead{(degree)} & \colhead{($z_s$)} & \colhead{(")} & \colhead{(AB)} & \colhead{(AB)} & \colhead{(AB)} & \colhead{(AB)} & \colhead{(AB)} & \colhead{(AB)}\\
\colhead{} & \colhead{(1)} & \colhead{(2)} & \colhead{(3)} & \colhead{(4)} & \colhead{(5)} & \colhead{(6)} & \colhead{(7)} & \colhead{(8)} & \colhead{(9)} &  \colhead{(10)} &  \colhead{(11)} &  \colhead{(12)} &  \colhead{(13)}
}
\startdata
1 & 000439.97-000146.4 & 1.16655    & -0.029549 & 1.166684   & -0.029906 & 0.583 & 1.23 & 20.61 & 20.42 & 20.06 & 23.53 & 21.88 & 21.0  \\
2 & 000508.37+010806.4 & 1.284914   & 1.13513   & 1.284664   & 1.134783  & 1.313 & 1.54 & 20.58 & 20.29 & 20.28 & 21.87 & 21.38 & 21.2  \\
3 & 000837.66-010313.7 & 2.156933   & -1.05378  & 2.156824   & -1.053306 & 1.37  & 1.75 & 20.91 & 20.52 & 20.57 & 21.02 & 20.7  & 20.66 \\
4 & 011935.29-002033.5 & 19.897043  & -0.342661 & 19.897005  & -0.342437 & 0.858 & 0.77 & 20.63 & 20.52 & 20.54 & 24.83 & 23.25 & 21.08 \\
5 & 012716.17-003557.6 & 21.817361  & -0.599367 & 21.817387  & -0.599763 & 0.371 & 1.5  & 21.87 & 20.45 & 19.94 & 23.04 & 21.43 & 20.73 \\
6 & 020318.87-062321.3 & 30.828754  & -6.389321 & 30.828445  & -6.389276 & 2.16  & 1.12 & 21.97 & 21.57 & 21.45 & 21.97 & 21.65 & 21.35 \\
7 & 021352.67-021129.4 & 33.469465  & -2.191534 & 33.468784  & -2.192038 & 2.78  & 2.99 & 20.91 & 20.69 & 20.78 & 21.1  & 20.67 & 20.85 \\
8 & 021930.51-055643.0 & 34.877115  & -5.94529  & 34.8773    & -5.945805 & 0.292 & 1.77 & 19.39 & 19.03 & 19.05 & 21.06 & 20.16 & 20.22 \\
9 & 022105.64-044101.5 & 35.273559  & -4.683761 & 35.273184  & -4.683806 & 0.199 & 1.23 & 19.85 & 18.75 & 18.38 & 20.01 & 18.76 & 18.48 \\
10 & 022159.71-014512.0 & 35.49886   & -1.753338 & 35.498105  & -1.753375 & 2.36  & 2.72 & 20.29 & 19.93 & 19.83 & 20.63 & 20.44 & 20.42 \\
11 & 084710.40-001302.6 & 131.793343 & -0.217354 & 131.793499 & -0.217577 & 0.627 & 0.95 & 19.75 & 19.32 & 18.95 & 20.17 & 19.91 & 19.41 \\
12 & 084856.08+011540.0 & 132.233475 & 1.260918  & 132.23375  & 1.260762  & 0.646 & 1.21 & 21.9  & 21.24 & 20.58 & 20.33 & 19.89 & 19.75 \\
13 & 090347.33+002026.2 & 135.947256 & 0.340628  & 135.946696 & 0.340352  & 0.412 & 2.22 & 18.98 & 18.88 & 18.82 & 21.73 & 20.46 & 19.43 \\
14 & 094132.90+000731.1 & 145.38711  & 0.125321  & 145.386881 & 0.125514  & 0.489 & 1.05 & 20.8  & 20.48 & 20.18 & 23.03 & 21.67 & 21.27 \\
15 & 121405.12+010205.1 & 183.521384 & 1.034736  & 183.521296 & 1.035337  & 0.493 & 2.18 & 20.15 & 20.07 & 19.8  & 20.15 & 19.96 & 19.7  \\
16 & 123821.66+010518.6 & 189.590277 & 1.088521  & 189.590554 & 1.088474  & 3.159 & 1.01 & 19.34 & 18.84 & 18.6  & 21.45 & 20.88 & 20.64 \\
17 & 141637.44+003352.2 & 214.156087 & 0.5645    & 214.155911 & 0.564565  & 0.434 & 0.66 & 20.09 & 19.94 & 19.66 & 21.16 & 20.35 & 19.64 \\
18 & 220228.65+004901.9 & 330.619373 & 0.817183  & 330.619666 & 0.817316  & 0.666 & 1.22 & 21.1  & 20.64 & 20.5  & 21.85 & 20.56 & 19.47 \\
19 & 220501.19+003122.8 & 331.254929 & 0.523039  & 331.25522  & 0.523039  & 1.652 & 1.07 & 20.53 & 20.31 & 20.13 & 21.23 & 20.95 & 20.88 \\
20 & 220718.43-001723.1 & 331.826821 & -0.289766 & 331.826869 & -0.289246 & 0.709 & 1.85 & 21.37 & 20.83 & 20.43 & 22.83 & 21.53 & 21.08 \\
21 & 220811.56+023830.1 & 332.048192 & 2.641706  & 332.047834 & 2.641156  & 3.179 & 2.36 & 21.44 & 21.16 & 21.11 & 20.46 & 20.13 & 20.02 \\
22 & 220906.91+004543.9 & 332.278805 & 0.762187  & 332.278736 & 0.761726  & 0.446 & 1.67 & 20.51 & 20.23 & 19.92 & 21.32 & 20.62 & 20.25 \\
23 & 221115.06-000030.9 & 332.81277  & -0.00859  & 332.812426 & -0.008998 & 0.478 & 1.9  & 19.74 & 19.84 & 19.75 & 22.4  & 21.16 & 20.56 \\
24 & 221227.74+005140.5 & 333.115601 & 0.861307  & 333.115689 & 0.860823  & 1.772 & 1.77 & 19.37 & 19.31 & 19.0  & 20.95 & 20.62 & 20.52 \\
25 & 221759.58+060426.3 & 334.498266 & 6.07399   & 334.498906 & 6.073757  & 2.572 & 2.44 & 19.83 & 19.43 & 19.17 & 21.43 & 20.21 & 19.64 \\
26 & 222057.44+000329.8 & 335.23933  & 0.058319  & 335.238828 & 0.058287  & 2.26  & 1.79 & 19.25 & 18.88 & 18.7  & 22.11 & 21.76 & 21.63 \\
27 & 222929.45+010438.4 & 337.372828 & 1.077209  & 337.372612 & 1.077473  & 1.445 & 1.21 & 21.25 & 20.89 & 20.73 & 22.48 & 21.25 & 20.83 \\
28 & 225147.82+001640.5 & 342.949299 & 0.277953  & 342.949571 & 0.277511  & 0.41  & 1.86 & 20.02 & 19.82 & 19.81 & 21.01 & 20.92 & 20.59 \\
29 & 230322.74-001438.2 & 345.844788 & -0.243969 & 345.844857 & -0.243347 & 3.222 & 2.25 & 20.61 & 19.89 & 19.74 & 21.41 & 21.09 & 21.01 \\
30 & 231152.90-001335.0 & 347.970424 & -0.226406 & 347.970578 & -0.226874 & 0.348 & 1.75 & 18.79 & 18.93 & 18.91 & 22.82 & 21.06 & 20.7  \\
31 & 232853.40+011221.8 & 352.222493 & 1.206087  & 352.222657 & 1.206152  & 0.528 & 0.66 & 21.81 & 21.24 & 20.31 & 22.85 & 21.76 & 20.92 \\
32 & 233713.66+005610.8 & 354.306952 & 0.936318  & 354.307046 & 0.936686  & 0.708 & 2.21 & 19.97 & 19.73 & 19.22 & 20.54 & 20.29 & 19.82 \\
\enddata
\tablecomments{\\
The basic information for the 32 targets we have observed.\\
Column (2) to Column (5): Coordinates of both sources measured with our decomposition methods.\\
Column (6): SDSS redshift of the known quasars.\\
Column (8) to Column (13): PSF magnitudes of both sources measured with our decomposition methods.\\
}
\end{deluxetable*}

\begin{deluxetable*}{l|lcccccccc}
\tablenum{2}
\tablecaption{Observation setups\label{tab:observation}}
\tablewidth{0pt}
\tablehead{
\colhead{} & \colhead{Name} & \colhead{Instrument} & \colhead{Grism+Filter} & \colhead{PA} & \colhead{$w_s$} & \colhead{Date} & \colhead{Seeing} & \colhead{Airmass} & \colhead{Exposure}\\
\colhead{} & \colhead{(J2000)} & \colhead{} & \colhead{} & \colhead{(degree)} & \colhead{(")} & \colhead{(dd.mm.yy)} & \colhead{(")} & \colhead{} & \colhead{(s)}\\
\colhead{} & \colhead{(1)} & \colhead{(2)} & \colhead{(3)} & \colhead{(4)} & \colhead{(5)} & \colhead{(6)} & \colhead{(7)} & \colhead{(8)} & \colhead{(9)}
}
\startdata
1 & 011935.29-002033.5 & Keck/LRIS & 600/7500; 600/4000 & -9.7 & 1 & 10.01.19 & 0.5 & 1.15 & $900; 880$\\
2 & 084710.40-001302.6 & Keck/LRIS & 600/7500; 600/4000 & -34.96 & 1 & 10.01.19 & 0.6 & 1.45 & $900; 880$\\
3 & 084856.08+011540.0 & Keck/LRIS & 600/7500; 600/4000 & -60.4 & 1 & 10.01.19 & 0.6 & 1.7 & $900; 880$\\
4 & 090347.33+002026.2 & Keck/LRIS & 600/7500; 600/4000 & 63.8 & 1 & 10.01.19 & 0.6 & 1.4 & $900; 880$\\
5 & 121405.12+010205.1 & Keck/LRIS & 600/7500; 600/4000 & -8.3 & 1 & 11.01.19 & 0.4 & 1.06 & $600; 590$\\
6 & 141637.44+003352.2 & Keck/LRIS & 600/7500; 600/4000 & -69.53 & 1 & 04.02.19 & 0.5 & 1.07 & $600; 580$\\
7 & 220228.65+004901.9 & Subaru/FOCAS & VPH850+O58 & 65.64 & 0.8 & 18.09.19 & 0.28 & 1.3 & $180\times3$\\
8 & 220718.43-001723.1 & Subaru/FOCAS & VPH850+O58 & 5.29 & 0.8 & 18.09.19 & 0.33 & 1.33 & $300\times3$\\
9 & 220906.91+004543.9 & Subaru/FOCAS & VPH850+O58 & 8.5 & 0.8 & 18.09.19 & 0.34 & 1.4 & $300\times3$\\
10 & 221115.06-000030.9 & Subaru/FOCAS & VPH850+O58 & 40.05 & 0.8 & 18.09.19 & 0.33 & 1.6 & $300\times1$\\
11 & 225147.82+001640.5 & Subaru/FOCAS & VPH850+O58 & -31.59 & 0.8 & 18.09.19 & 0.38 & 1.47 & $600\times2$\\
12 & 231152.90-001335.0 & Subaru/FOCAS & VPH850+O58 & -18.18 & 0.8 & 18.09.19 & 0.32 & 1.54 & $600\times2$\\
13 & 232853.40+011221.8 & Subaru/FOCAS & VPH850+O58 & 68.61 & 0.8 & 18.09.19 & 0.27 & 1.59 & $600\times2$\\
14 & 000439.97-000146.4 & Subaru/FOCAS & VPH850+O58 & -20.5 & 0.8 & 18.09.19 & 0.3 & 1.54 & $600\times2$\\
15 & 021930.51-055643.0 & Subaru/FOCAS & VPH850+O58 & -19.68 & 0.8 & 18.09.19 & 0.25 & 1.2 & $300\times2$\\
16 & 220501.19+003122.8 & Subaru/FOCAS & 300B+Y47 & -90.15 & 0.8 & 19.09.19 & 0.44 & 1.25 & $600\times2$\\
17 & 220811.56+023830.1 & Subaru/FOCAS & 300B+Y47 & 33.03 & 0.8 & 19.09.19 & 0.22 & 1.37 & $600\times2$\\
18 & 221759.58+060426.3 & Subaru/FOCAS & 300B+Y47 & -69.9 & 0.8 & 19.09.19 & 0.27 & 1.43 & $600\times1$\\
19 & 222057.44+000329.8 & Subaru/FOCAS & 300B+Y47 & -93.64 & 0.8 & 19.09.19 & 0.31 & 1.65 & $600\times1$\\
20 & 221227.74+005140.5 & Subaru/FOCAS & 300B+Y47 & -10.33 & 0.8 & 19.09.19 & 0.32 & 1.9 & $600\times1$\\
21 & 230322.74-001438.2 & Subaru/FOCAS & 300B+Y47 & 6.32 & 0.8 & 19.09.19 & 0.29 & 1.62 & $600\times2$\\
22 & 233713.66+005610.8 & Subaru/FOCAS & 300B+Y47 & 14.32 & 0.8 & 19.09.19 & 0.38 & 1.54 & $600\times2$\\
23 & 000508.37+010806.4 & Subaru/FOCAS & 300B+Y47 & 35.74 & 0.8 & 19.09.19 & 0.33 & 1.54 & $600\times2$\\
24 & 000837.66-010313.7 & Subaru/FOCAS & 300B+Y47 & -13.02 & 0.8 & 19.09.19 & 0.62 & 1.74 & $600\times1$\\
25 & 021352.67-021129.4 & Subaru/FOCAS & 300B+Y47 & 53.5 & 0.8 & 19.09.19 & 0.45 & 1.21 & $600\times2$\\
26 & 022159.71-014512.0 & Subaru/FOCAS & 300B+Y47 & -92.81 & 0.8 & 19.09.19 & 0.88 & 1.31 & $600\times1$\\
27 & 233713.66+005610.8 & Gemini/GMOS-N & R831+RG610 & 14.32 & 0.75 & 08.09.19 & 0.5 & 1.13 & $605\times3$\\
28 & 012716.17-003557.6 & Gemini/GMOS-N & R831+RG610 & -3.87 & 0.75 & 08.09.19 & 0.5 & 1.42 & $605\times2$\\
29 & 020318.87-062321.3 & Gemini/GMOS-N & R400+OG515 & -81.65 & 0.75 & 08.09.19 & 0.4 & 1.12 & $840\times3$\\
30 & 220501.19+003122.8 & Gemini/GMOS-N & R831+OG515 & -90.15 & 0.75 & 24.09.19 & 0.5 & 1.19 & $605\times3$\\
31 & 220228.65+004901.9 & Gemini/GMOS-N & R831+RG610 & 65.64 & 0.75 & 24.09.19 & 0.45 & 1.39 & $605\times3$\\
32 & 222929.45+010438.4 & Gemini/GMOS-N & R831+OG515 & -39.28 & 0.75 & 25.09.19 & 0.3 & 1.17 & $605\times3$\\
33 & 022105.64-044101.5 & Gemini/GMOS-N & R831+OG515 & 83.13 & 0.75 & 25.09.19 & 0.3 & 1.14 & $225\times4$\\
34 & 094132.90+000731.1 & Gemini/GMOS-N & R831+OG515 & -49.81 & 0.75 & 19.12.19 & 0.6 & 1.07 & $605\times4$\\
35 & 123821.66+010518.6 & Gemini/GMOS-N & B600+CG455 & -80.34 & 0.75 & 24.12.19 & 0.4 & 1.13 & $847\times2$\\
\enddata
\tablecomments{\\
Observational setups of the targets, rearranged by observation date.\\
Column (5): The widths of the long slits.\\
Column (3) and (9): For LRIS, data are taken with two cameras simultaneously, left of the semicolon is the setting for red camera, right for the blue.\\
No.7 \& No.31, No.16 \& No.30, No.22 \& No.27: These targets were observed twice with different instruments for reasons of wavelength coverage.
}
\end{deluxetable*}

\subsection{Subaru/FOCAS}\label{subsec:focas}
We observed twenty candidates on 18.09.2019 and 19.09.2019 using the Faint Object Camera and Spectrograph (FOCAS: \cite{10.1093/pasj/54.6.819}) on the 8.2m Subaru telescope (S19B-079, PI: J. Silverman). Nine targets were observed on the first night with the grism VPH850+O58, which covers the wavelength range from 5800~\text{\AA} to 10350~\text{\AA} and has a spectral dispersion of 1.17~\text{\AA}/pixel. During the second night, we observed eleven targets with the 300B+Y47 grating, that covers wavelengths from 4700 to 9100~\text{\AA} with a dispersion of  1.34~\text{\AA}/pixel. The slit width was $0^{\prime\prime}.8$ for both nights. Detailed information is listed in Table \ref{tab:observation}.\par
Data reduction was carried out with the \textbf{focasred} package, following the Subaru Data Reduction CookBook\footnote{\url{https://www.naoj.org/Observing/DataReduction/Cookbooks/FOCAS_cookbook_2008dec08.pdf}}. We started with bias subtraction and flat field correction. We did not correct for the spatial distortion to avoid introducing additional noise. Then we applied L.A. COSMIC, which is a cosmic ray removal tool based on a variation of Laplacian edge detection developed by \cite{van2001cosmic} to remove cosmic ray events. Subsequently, we implemented the IRAF \textbf{apall} task in \textbf{apextract} package to extract the trace of the 1D spectrum. Wavelength calibration was performed 
using ThAr lamp exposures. Finally, we observed the standard star G191B2B with the same setup as the science exposures for flux calibration.\par

\subsection{Gemini/GMOS-N}
We observed nine candidates with the Gemini Multi-Object Spectrograph (GMOS; \cite{hook2004gemini}) on the 8.1m Gemini-North telescope (GN-2019B-Q-128, PI: J Silverman) from September 2019 to December 2019 in the queue mode. Depending on the redshift of the dual quasar candidate, a different setup including the choice of grating was configured to cover a wavelength range suitable to detect broad emission lines (e.g. C\,{\sc iv}, Mg\,{\sc ii}, H$\alpha$) characteristic of quasar activity (Table \ref{tab:basic}).\par
The data was taken with the GMOS-N Hamamatsu detectors. CuAr lines were used for wavelength calibration. The observed standard stars were EG131 and Feige 66. The reduction was done using the semi-automated Python Spectroscopic Data Reduction Pipeline \citep[PypeIt ][]{Prochaska2020}. Parameters were tuned according to the documents to fit the usage of each object.
\par
Considering the different algorithms used in IRAF and PypeIt, we normalize the results by rescaling the flux calibrated spectra to match the PSF magnitudes of the sources measured from our image decomposition. These HSC photometric data points are plotted as star symbols in panel b of Figures \ref{fig:2209} to \ref{fig:2311}.

\subsection{Spectral Classification}
To confirm a dual quasar system, we classify the companion based on its optical spectrum. Generally, when broad emission lines such as C\,{\sc iv}, CIII], Mg II, H$\beta$ or H$\alpha$ are observed, we classify the companion as a type-1 quasar, while diagnostic diagrams (``Baldwin, Phillips \& Terlevich" diagrams \cite{baldwin1981classification} and WHAN diagram \cite{cid2011comprehensive}) are used to classify type-2-like sources. Broad absorption lines are also common among quasars (e.g. \cite{hewett2003frequency}); for example, C\,{\sc iv} may form a so-called "P-Cygni" profile, with a broad absorption feature impacting the blue wing of the core emission. There is one such case for which we classify as a ``BAL quasar". Table \ref{tab:duals} lists the six dual quasars that we have confirmed to date. Classification of the rest of the targets are reported in Appendix \ref{sec:by_products}.

\begin{deluxetable*}{l|llccl}
\tablenum{3}
\tablecaption{Confirmed dual quasars\label{tab:duals}}
\tablewidth{0pt}
\tablehead{
\colhead{} & \colhead{Name} & \colhead{Type} & \colhead{$z_A$} & \colhead{$z_B$} & \colhead{Features}\\
\colhead{} & \colhead{(1)} & \colhead{(2)} & \colhead{(3)} & \colhead{(4)} & \colhead{(5)}
}
\startdata
\hline
1 & 084710.40-001302.6$^*$ & type 1 Quasar & 0.6255 & 0.6268 & broad Mg\,{\sc ii}\\
2 & 121405.12+010205.1$^*$ & type 1 Quasar & 0.4916 & 0.4935 & broad H$\alpha$ and H$\beta$\\
3 & 123821.66+010518.6 & BAL Quasar   & 3.1143 & 3.1268 & C\,{\sc iv} BAL\\
4 & 141637.44+003352.2$^*$ & type 2 Quasar & 0.4331 & 0.4327 & [Ne\,{\sc v}], [O\,{\sc ii}], [Ne\,{\sc iii}], and [O\,{\sc iii}] emission\\
5 & 220906.91+004543.9 & type 1 Quasar & 0.4457 & 0.4457 & broad H$\alpha$ and H$\beta$\\
6 & 233713.66+005610.8 & type 1 Quasar & 0.7089 & 0.7083 & broad H$\beta$, [O\,{\sc iii}] outflow and iron emission\\
\hline
\enddata
\tablecomments{\\
Spectroscopically confirmed physically associated quasar pairs.\\
Column (2): Type of the companion quasar. The main quasars are always type 1.\\
Column (5): Notable observed features.\\
$^*$ : From our previous work \citep{silverman2020dual}.
}
\end{deluxetable*}

\subsection{Completeness}
\begin{figure}[htp]
\centering
\includegraphics[width=0.5\textwidth]{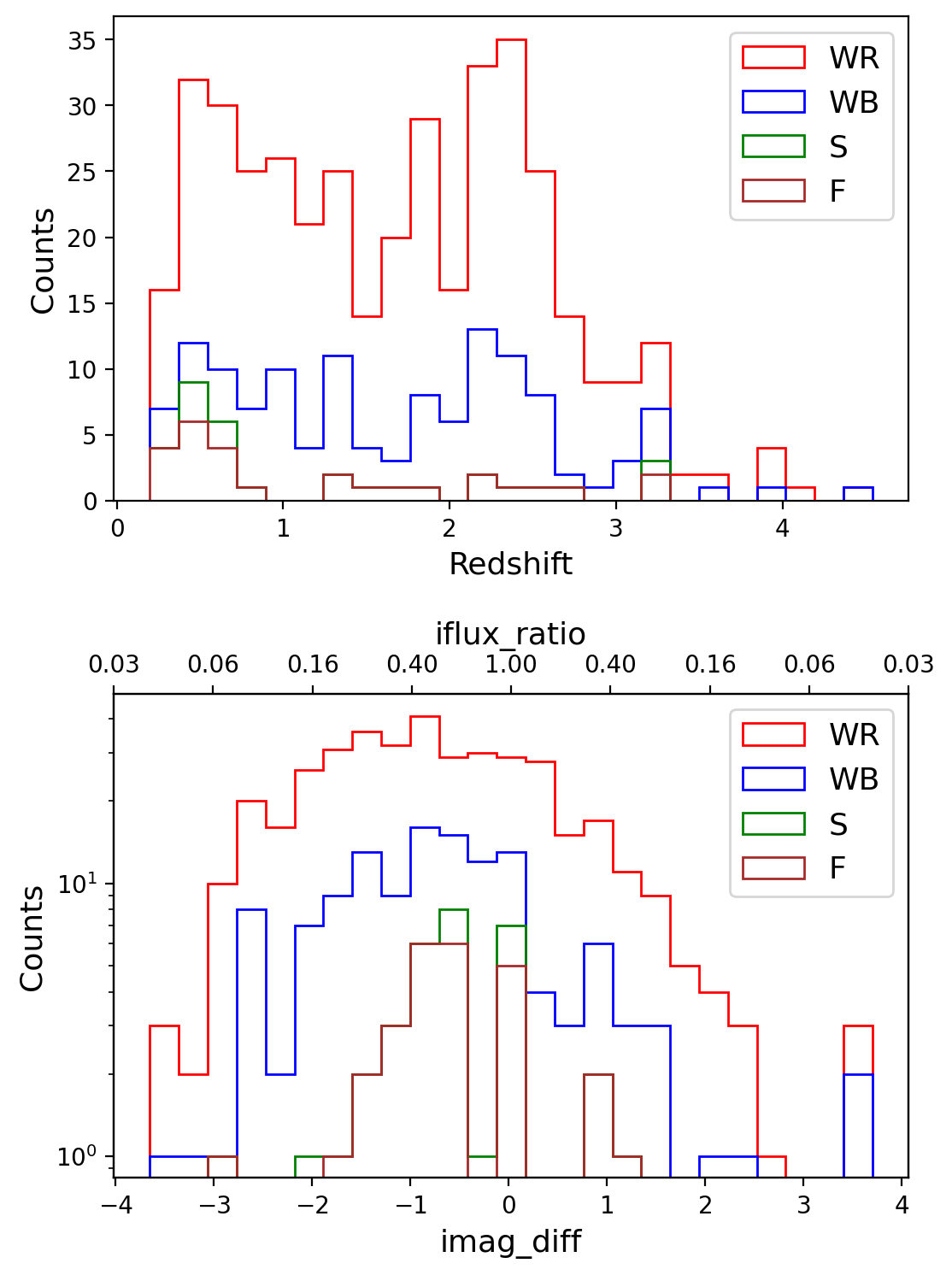}
\caption{Target completeness with respect to redshift (upper panel) and $i$-band luminosity and magnitude difference (bottom panel, $quasar\ imag - companion\ imag$ on the bottom axis, and the corresponding flux ratio on the upper axis). The histograms are stacked with the red and blue components showing the candidates that have not been observed (``W"ait, ``R"ed for those with $g-r>1$, and ``B"lue for those with $g-r<=1$), brown components are "F"ailure cases (i.e., the companion is not a physically associated quasar), and green components are the ``S"uccess cases, i.e the companions are spectroscopically confirmed as physically associated quasars.}
\label{fig:z_success}
\end{figure}

We show the completeness of this work in Figure \ref{fig:z_success} with stacked histograms binned by redshift and $i$ band magnitude difference. The colors refer to the status of the targets, either observed (brown for the failures and green for the successes) or not observed (separated by the companions' $g-r$ color). The upper panel gives a statistical view of Figure \ref{fig:bol_z} on the x-axis. Our follow-up observations covered redshift from 0 to $\sim$3.2, including 20 blue targets and 12 red targets. The $i$ band magnitude differences of the pairs correspond to flux ratios from 0.03 to 1. Therefore, our selection algorithm is able to cover pairs that differ in flux by up to a factor of $\sim$ 30.  

\section{Emission-line fitting}\label{sec:Fitting}
\begin{figure*}[htb]
\includegraphics[width=170mm]{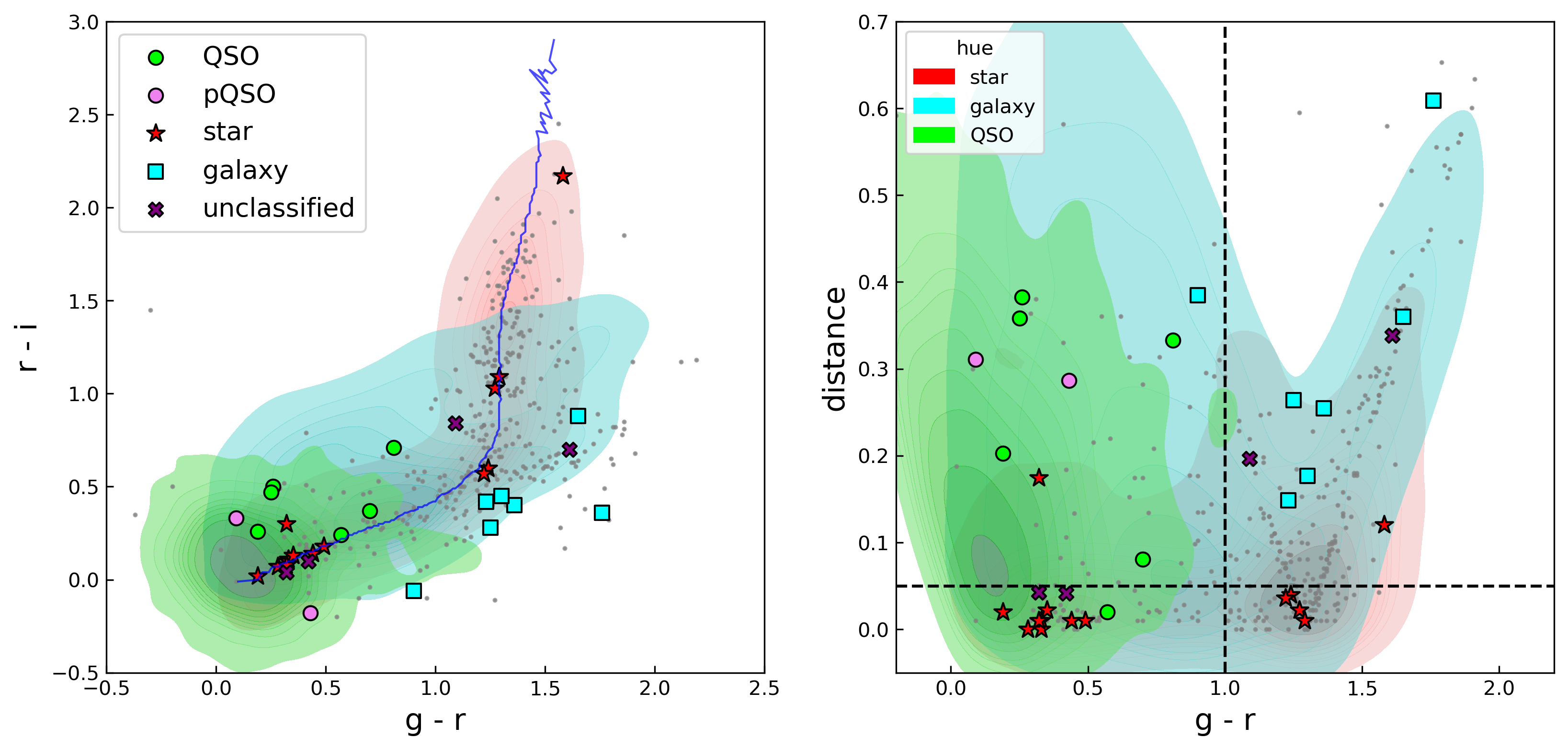}
\caption{Distribution of the companions in color space. Left panel: color distribution around the stellar locus. Green dots are quasars physically associated with the main quasar (not shown). Pink dots are projected quasars. Stars are marked in red. Galaxies are indicated by the cyan squares. Purple crosses mark unclassified sources. Right panel: the distance of each target from the stellar locus is shown as a function of their $g-r$ color. Dashed lines highlight values of $g-r=1.04$ and $distance=0.06$, which we found to be a reasonably effective classification boundary. The background shadows show the KDE plot of $\sim 4000$ stars, $\sim 4000$ galaxies, and $\sim 4000$ quasars selected from the same HSC PDR2 catalog. The default linear normalization is used for the contour levels.
\label{fig:distance}}
\end{figure*}
As mentioned in Section \ref{sec:Spec}, a line width is required to confirm a type 1 AGN while the line ratio is used in the BPT diagram to confirm a type 2 AGN. Therefore, we fit the spectrum of AGN candidates to decompose them into continuum and emission lines and estimate the widths and heights of the lines. We use the \textbf{scipy.optimize.curve\_fit} package in Python to perform the fitting. The optimization is based on the Trust Region Reflective algorithm \citep{POWELL197031}. Given the least square loss function, the basic idea of the algorithm is to approximate it with a simpler function, which reflects the behavior in its trust region. Each step is taken to minimize the loss function over the trust region. In most cases, we set some constraints on the parameters to have convergence.
\par
Here we describe the models applied to specific emission lines. For C\,{\sc iv} $\lambda$1549, we first fit a power law in the spectral window [1445,1465] \text{\AA} and [1700,1705] \text{\AA}, to remove the continuum. We fit the emission line with either one or two Gaussian components. For each Gaussian component, the free parameters are its centroid, height, and width. The parameters for the two Gaussians are not tied together. 
We mask regions that are affected by significant absorption features. If a source has BAL features, we include an extra Gaussian component to describe the absorption feature in the model fit. Similar to C\,{\sc iv} $\lambda$1549, C\,{\sc iii}] $\lambda$1909 and Mg\,{\sc ii} $\lambda$2798 are also usually single component lines, thus a single Gaussian component after subtracting the continuum is sufficient in most cases. The power law continuum region for C\,{\sc iii}] $\lambda$1909 is measured between 1700 and 1820 \text{\AA}; 2000 and 2500 \text{\AA}. For Mg\,{\sc ii} $\lambda$2798, it is measured between 2200 and 2700 \text{\AA}; 2900 and 3090 \text{\AA}. For the case of Mg\,{\sc ii}, iron emission is sometimes significant, thus we implemented the \cite{vestergaard2001empirical}'s UV iron template as an additional component when necessary.
\par
H$\beta$ and [O\,{\sc iii}] $\lambda\lambda$4959, 5007 are fitted simultaneously. The spectral windows [4435, 4700] \text{\AA} and [5100, 5535] \text{\AA} are used to fit the continuum and the iron template is applied when we find it is necessary to improve the results. After subtracting the continuum and iron emission, we fit the emission line components together in several cases depending on our data: a typical case includes three Gaussian components, corresponding to a narrow H$\beta$ component + [O\,{\sc iii}] $\lambda$4959 + [O\,{\sc iii}] $\lambda$5007. Unlike the cases of C\,{\sc iv} and Mg\,{\sc ii}, here we set bounds between the parameters of the different Gaussian components. We assume the widths of the narrow lines to be the same in velocity space and locked their centroid position, so that the separation between the centroid of [O\,{\sc iii}] $\lambda$4959 and that of [O\,{\sc iii}] $\lambda$5007 is always 48~\text{\AA} in the rest frame, and the separation between H$\beta$ and [O\,{\sc iii}] $\lambda$5007 is always 145.5~\text{\AA}). We also set the ratio of the [O\,{\sc iii}] doublet's height to be the atomic physics value of 2.98  \citep{storey2000theoretical}. Therefore, in this case, we only have one free positional parameter, one free width parameter and two free height parameters. When a broad component is required for H$\beta$, we add one extra Gaussian component to the model, whose  parameters are independent of the narrow components.
On the other hand, it is common for the [O\,{\sc iii}] doublet to have asymmetric blue wings \citep{greene2005comparison}; in that case, we add a Gaussian component located blueward of the core components, which has the same width, proportional height and locked position in the two lines of the doublet.
\par
H$\alpha$, [N\,{\sc ii}] $\lambda\lambda$6548,6584 and [SII] $\lambda\lambda$6716,6731 are fitted simultaneously. The continuum is measured between 6000 and 6250 \text{\AA}, and 6800 and 7000~\text{\AA}. Similar to the modelling of H$\beta$, the basic case includes five Gaussian components with each of them corresponding to a narrow emission line sharing the same width. The ratio of [N\,{\sc ii}] $\lambda$6584 to [N\,{\sc ii}] $\lambda$6548 is set to 2.96, while the heights of the two lines of the [SII] doublet are free. When a broad component is required for H$\alpha$, we add one extra Gaussian component to the model.

\section{Results} \label{sec:Results}
We first report on the fraction of our spectra that reveal a true dual quasar (the success rate), and its dependence on color. We then describe the properties of three newly discovered quasar pairs, two projected quasar pairs, and three quasar-galaxy pairs from this study. These newly identified dual quasars will improve the sample that allows us to investigate the relation between SMBH mass and that of their host galaxies at $z < 1$ for the first time in merging systems (Section \ref{sec:Discussion}). In our following discussions about colors, we always refer to the HSC color; the SDSS color is calibrated to the HSC system as described in Appendix \ref{sec:Calibration}. 
\subsection{Spectroscopic Success Rate and Dependence on Optical Properties}\label{subsec:Rate}

\begin{figure*}[htp]
\begin{centering}{}
\includegraphics[width=150mm]{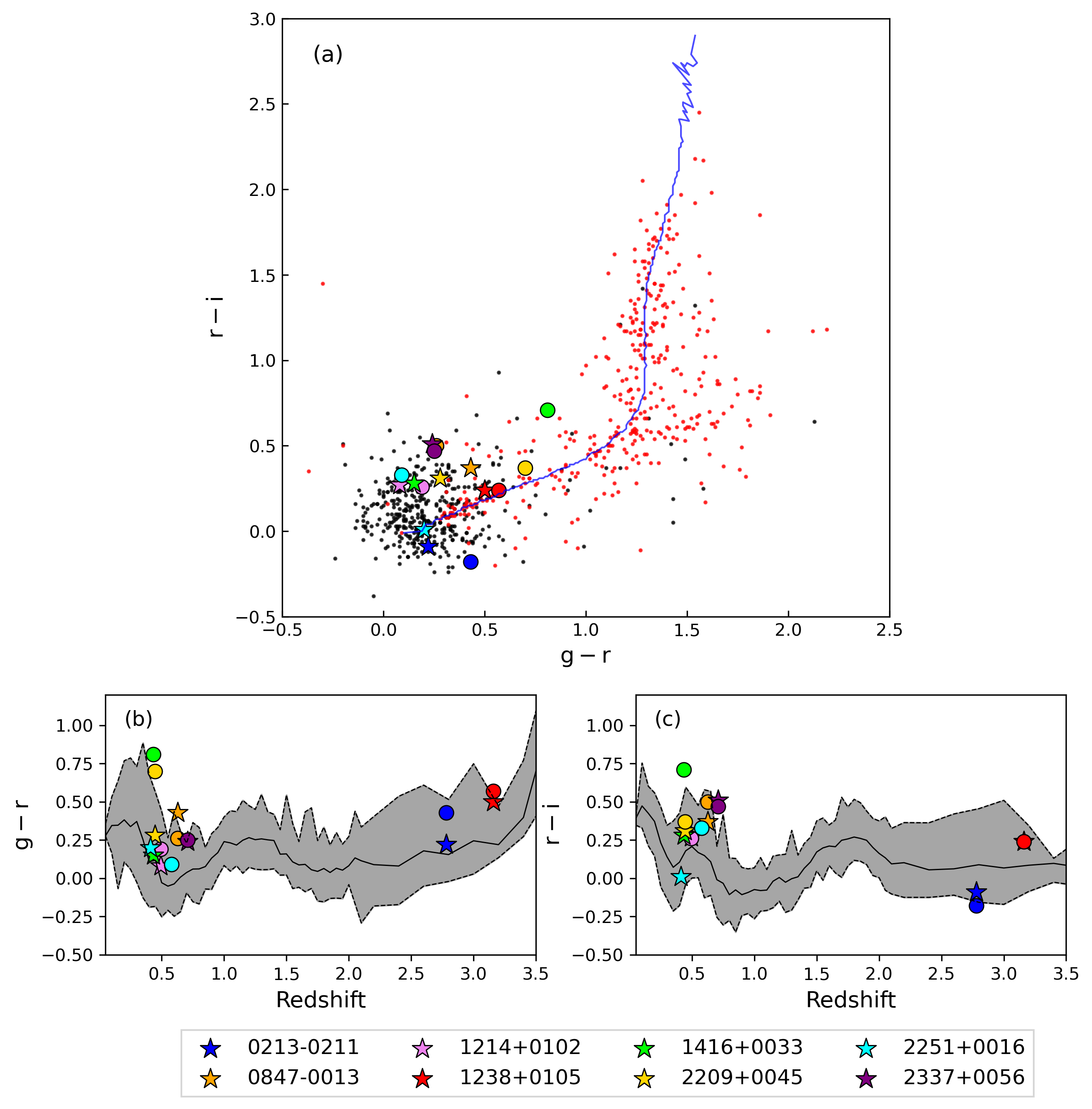}
\caption{Quasar pairs in color space. The upper panel is similar to Figure \ref{fig:distance} (left panel). Here, quasars and their companions are shown (black dots: main SDSS quasars; red dots: companions). The blue line is the stellar locus \citep{covey2007stellar}. Bottom panels show the quasar pairs in color--redshift space. The average quasar color \citep{richards2001colors} is shown by the black curve with the grey shadow representing the 95\% percentile regions for the corresponding colors. Quasar pairs are color-coded with the star symbols representing the primary source and the circular symbols representing the companions.
\label{fig:SDSS_color}}
\end{centering}
\end{figure*}

Out of the 32 targets we followed up spectroscopically, we confirm six quasar pairs, giving a success rate of $\sim 19\%$ for finding physically associated quasar pairs. This success rate is significantly lowered by the inclusion of candidates with colors in agreement with the stellar locus, which are necessary to cover those red quasars. Below, we discuss means of limiting our rate of failure. In addition, we identified three AGN-galaxy pairs (only one SMBH in the pair is active) and two project quasars pairs in which the quasars are at significantly different redshift (velocity offset larger than 2000 $\mathrm{km\,s^{-1}}$).
\par

Figure \ref{fig:distance} provides a view of these results in color--color space, with only the previously unknown sources plotted. In the left panel, the blue curve is a stellar locus adopted from \cite{covey2007stellar}, which shows the average distribution of stars as a reference. The grey dots are our 401 candidates. The magnitudes used here are the PSF magnitudes estimated from our decomposition methods based on HSC imaging \citep{ding2020mass, li2021sizes}. We claim that the measurements of AGN magnitudes are precise since the light distribution of the point sources are quite unique, the uncertainties are under 0.02 mag within a certain model (e.g. 2 Point Sources + 1 S\'ersic Profile). In some cases, the spectrum is too noisy or too featureless to tell which class they belong to; these unclassified targets are labeled with a purple cross. 
\par
For comparison, we show the kernel density estimation (KDE) plot for normal HSC stars, galaxies, and quasars respectively. The galaxies are selected with $r_{extendness} = 1$ and $i_{extendness} = 1$, which means the HSC Pipeline \citep{bosch2018hyper} classified them as extended sources in both the $r$ and $i$ bands. Stars are selected as point sources in these two bands with $r_{extendness} = 0$ and $i_{extendness} = 0$. While this sample will include some quasars, stars outnumber quasars by several orders of magnitude, and it is safe to consider it as a statistical star sample. Quasars are selected from our parent sample pool, i.e., the HSC imaged SDSS quasars (Section \ref{subsec:Candidates}). We reject objects with $g-r$ outside the $1-99\%$ range of the distribution for the stars, galaxies, and quasars, to remove objects with poor photometry. 
Finally, $\sim 4000$ targets are randomly selected from each of the categories as the comparison sample for Figure \ref{fig:distance}.
\par
We find that the classification of our candidates based on spectroscopy is consistent with the KDE distribution of the comparison categories. Since the comparison quasar sample is from SDSS, they are mostly type 1 quasars, and thus are concentrated at the lower left region of the $r-i$ vs $g-r$ plot. As expected, this is also the region in color space where we have a high success rate. However, contamination from galaxies and stars is still significant. We represent the figure as a distance--color space (Figure \ref{fig:distance} right panel), where distance is defined as the offset of our targets from the stellar locus in the $r-i$ vs $g-r$ space. This separates the different populations to a certain extent. 
The quasars are systematically further away from the stellar locus than the stars, and are bluer than the galaxies. We draw a vertical dashed line at $g-r=1.0$ and a horizontal line at ``distance from stellar locus=0.06'', dividing the figure into four regions. We have a high success rate of identifying quasars in the upper left region(7 out of 9). One of the contaminants is a blue star, while another one is a narrow-line galaxy with emission line ratios that suggest it is a composite AGN and star forming object. 
\par
The lower region mostly contains stars, since they indicate a very close distance to the stellar locus. They are the main contaminants in the sample. While we might have expected to find type 2 companion quasars in this region of the diagram, 
the only type 2 quasar companion we found is SDSS J1416+0033 in the upper left region, although far from the bulk of the quasars with $g-r \sim 0.71$. One true quasar pair lies in the lower region is a pair of quasars at $z\sim3.1$. Its host galaxy is undetected,  making the source star-like, and it lies very close to the stellar locus. 
\par
Galaxies are more widely distributed in this diagram as shown by the comparison samples. The diffused galaxies are filtered out by our selection algorithm, left those with luminous bulges or star formation regions that might appear as fake point sources.
\par
\cite{hennawi2010binary} studied the colors of dual quasars to $z\sim5.5$ (see their Figure 4). High redshift quasars ($z>3$) are more scattered in the color-color space than the low-redshift quasars. And due to their relatively low number density, they are not recognizable in the contours of Figure \ref{fig:distance}. One has to consider redshift as well as the color when pursuing high redshift quasars.
\par
We now focus on the color of the quasar pairs. We plot both sources of each pair on Figure \ref{fig:SDSS_color}. In the upper panel, red dots are the same data set as in Figure \ref{fig:distance}, and black dots are the parent sample of these systems, i.e., the known SDSS quasars. Our eight quasar pairs are shown with star symbols, where the color indicates the two sources in a given system.  In particular, the blue and cyan star symbols denote the two projected quasar pairs (SDSS J0213-0221 and J2251+0016).
\par
\cite{richards2001colors} show that the locus of quasar colors as a function of redshift is very tight.  
The lower panels plot this locus; the shadowed region contains 95\% of the sample at any given redshift.  We 
overlay our physically-associated and projected quasar pairs. Eight of the 12 quasars in physically associated pairs fall in the 95\% region of $g - r$, 9 out of 12 fall in the 95\% region of $r - i$, but most of our objects lie above the mean color. Thus dual quasars have colors that are not drawn exclusively from the color distribution of SDSS-selected quasars.  The fact that they tend to be redder suggests that they are affected by dust in their hosts, due to the interaction of the two objects \citep{richards2003colors}.
\par
The two projected quasar pairs are not interacting with one another, suggesting that they should lie in the 95\% region. This is true for J2251+0016, but not for J0213-0211. The redshift separation between the two objects in the latter source is 6000 km s$^{-1}$, so it is conceivable that they are physically associated with one another. 
\par
It is also worth mentioning that the redshifts themselves of the sources are not precisely determined in all cases: the emission lines are broad and often asymmetric, so this can affect the measurement of the apparent line-of-sight velocity offset. The typical shift of C\,{\sc iv} is $\sim$ 800 km s$^{-1}$ compared to Mg\,{\sc ii}, while the later usually only have a small blueshift compared to [O\,{\sc iii}] \citep{richards2002broad}.
\par
In the following section, we report the details of newly-identified three physically-associated quasar pairs. The projected pairs are presented in \ref{sec:Projected}, and the quasar-galaxy pairs in \ref{sec:Galaxy}.

\subsection{Physically-associated quasar pairs}
We report on the properties of three newly-discovered dual quasars that are physically associated, which is defined as a quasar pair with line-of-sight velocity offset smaller than 2000 $\mathrm{km\,s^{-1}}$  \citep{hennawi2006binary}. The lens hypothesis is ruled out in each case by the fact that the spectra of the two objects in each pair are not identical.
\subsubsection{SDSS J220906.91+004543.9}

\begin{figure*}[htp]
\begin{centering}
\includegraphics[width=1\textwidth]{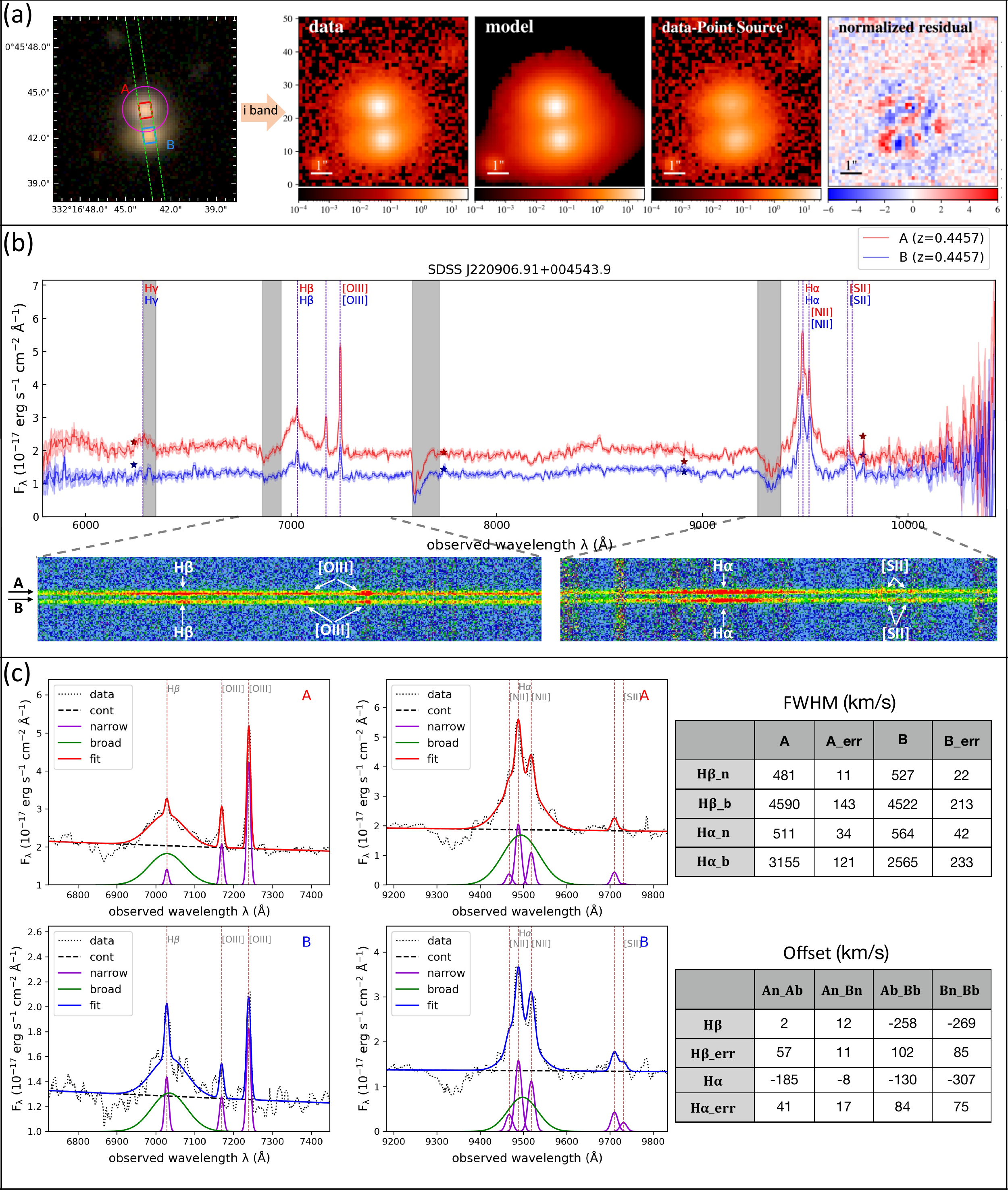}
\caption{A kaleidoscope of SDSS J2209+0045. Panel (a): HSC imaging overlaid with spectroscopic setups and decomposition results on $i$ band image. Panel (b): 1D and 2D spectrum, the 1D spectrum of each source is rescaled to match with the PSF magnitudes labeled with star symbols respectively. The grey shadows indicate the telluric absorption bands. Panel (c): Supplementary information. The upper row shows the line fitting results of source A, and the lower row shows source B. The tables show the FWHM and offset of the Balmer lines.
\label{fig:2209}}
\end{centering}
\end{figure*}

The SDSS survey reported this object as a broad-line quasar at $z = 0.4466$. The Subaru HSC images are shown in Figure \ref{fig:2209}a. The color image (RGB), shown in the left panel, is generated using the $g$, $r$ and $i$ bands \citep{lupton2004preparing}. The magenta circle shows the position of the SDSS fiber, whose diameter is 3\arcsec~for this target. The green dashed rectangular region marks the position of the long slit used for spectroscopy. The blue and red rectangular regions are the extraction apertures on the 2D spectrum for each source. The four figures on the right hand side show the image decomposition results using the \hbox{$i$-band} image and Lenstronomy-based algorithm \citep{birrer2018lenstronomy}. From left to right, the panels are the following: (1) original HSC data, (2) model reconstructed with point sources + host galaxies, (3) host galaxies of the system (data $-$ point source models),and (4) residual of data $-$ model normalized by the pixel errors. According to the decomposition results, the \hbox{$i$-band} magnitudes of the point sources are 19.92 and 20.25, respectively. We carry out a similar analysis for all five bands and find that the southern source is redder ($g-r=0.7$) than the typical SDSS quasar (Figure \ref{fig:SDSS_color}). The same is true for the primary quasar. We also measured the separation of the two point sources.
The point sources are separated by 1\arcsec.67, which corresponds to 9.55 kpc at $z = 0.4466$.
\par
The optical spectra are presented in the middle panels of Figure \ref{fig:2209}b. The 1D spectra are extracted from the rectangular regions shown in the HSC color image. We corrected the flux calibrated spectrum by matching to the HSC photometry, which are plotted as stars, dark red for source A and dark blue for source B. The position of emission lines are also labeled with dashed vertical lines having the same coding. The regions affected by telluric oxygen and water vapor absorption are shaded in grey (oxygen bands: 6280--6340 \text{\AA}, 6860--6950 \text{\AA}, 7590--7720 \text{\AA}; water vapor bands: 9270--9380 \text{\AA}). We show a zoom-in view of the 2D spectrum with the corresponding sources (A and B) labeled on the left. The traces are well-separated. The FOCAS spectrum of this target covers the H$\beta$ and H$\alpha$ regions.  
\par
The bottom panel (Figure \ref{fig:2209}c) provides supplemental and detailed information of the pair. Line-fitting results of the two sources are given with the first (second) row corresponding to source A (B). We fit the H$\beta$ and H$\alpha$ regions using the methods described in Section \ref{sec:Fitting}. Vertical dashed lines indicate the centroid of each emission line. The tables on the right-hand side record the widths of narrow and broad lines with errors. The offsets of the line centroids between the two sources or between the broad and narrow components of the same line are listed. For example, $``\mathrm{An\_Ab}''$ stands for the offset between the narrow line and the broad line of source A. $``\mathrm{Ab\_Bb}''$ stands for the offset between the broad line of source A and the same broad line of source B. The velocity is positive when the former line is redshifted relative to the latter line.
For SDSS J2209+0045, the measurement of the narrow lines indicate very small offsets, and both sources have broad emission lines $> 2000\,\mathrm{km\,s^{-1}}$. The difference of the line ratios tells us that this is not a lens system, and we thus classify it as a quasar pair. 
\par
In panel (a), the decomposition of the $i$-band image resolves the host galaxies of this system. The model indicates a Sérsic index of 0.8 for the main quasar and 3.8 for the companion. After subtracting the two Sérsic profiles and point-source components from the science image, the residual image seems to indicate an asymmetry of the galaxies' morphology, possibly resulting from the on-going merger event. However, for SDSS J2209+0045, the asymmetry is not very prominent hence the galaxies are not strongly distorted. Thus we suggest that this merging system is undergoing a first passage.

\subsubsection{SDSS J233713.66+005610.8}
\begin{figure*}[htp]
\begin{centering}
\includegraphics[width=1\textwidth]{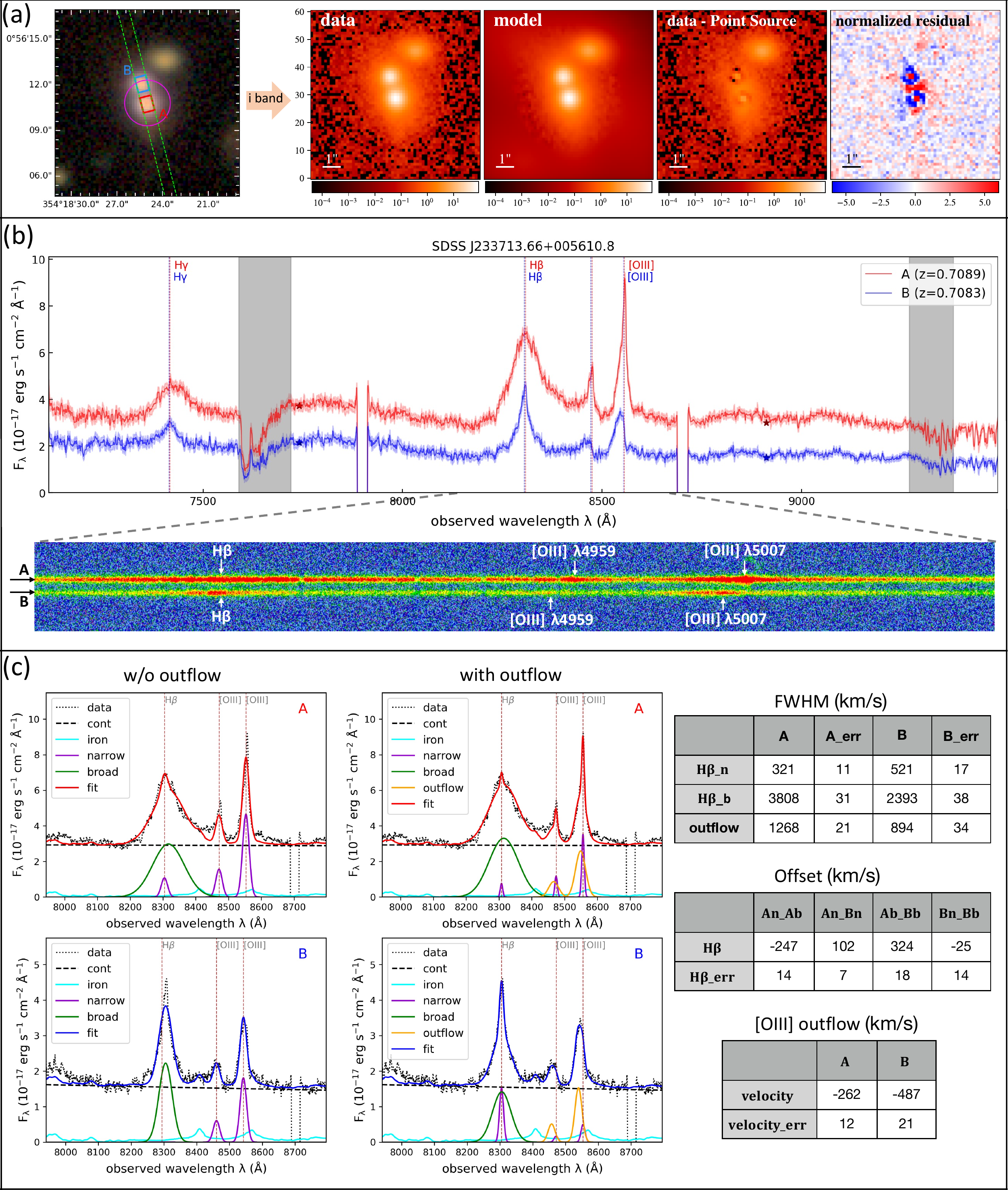}
\caption{A kaleidoscope of SDSS J2337+0056. Panel (a): HSC imaging overlaid with spectroscopic setups and decomposition results on $i$ band image. Panel (b): 1D and 2D spectrum, the 1D spectrum of each source is rescaled to match with the PSF magnitudes labeled with star symbols respectively. The grey shadows indicate the telluric absorption bands. Panel (c): Supplementary information. The left column of images show the line fitting results without outflow components, and the right column of images include the outflow components. The tables show the FWHM and offset of the Balmer lines, and the outflow velocity of [O\,{\sc iii}].
\label{fig:2337}}
\end{centering}
\end{figure*}

SDSS classified this object as a broad-line quasar at $z = 0.7080$. The fiber is centered on the southern source as shown in Figure \ref{fig:2337}a. The color of the two sources are very similar and both of them are located in the typical quasar region of the color-color diagram (Figure \ref{fig:SDSS_color}). The HSC image also shows a third source to the north of the pair, thus we fit 2 PSFs + 3 Sersic profiles for this system. We see the third source is well-modeled by the Sérsic profile and has no obvious features related to the pair. This suggests that this source is an independent foreground or background source.
\par
The GMOS spectrum of this target was taken with the R831+RG610 grism, and the H$\beta$ region was placed at the center of the spectrum (Figure \ref{fig:2337}b). The traces of the two objects in the 2D spectrum are well-separated and we can clearly see the [O\,{\sc iii}] and H$\beta$ features in both spectra. Notably, the [O\,{\sc iii}] doublet appears to have blue wings in both sources. To support this statement, we fit the lines with two different models (panel c). In the left column, no outflow components are used, thus each [O\,{\sc iii}] line is modelled with a single Gaussian. This leaves a significant residual at the peak of [O\,{\sc iii}] in source A and H$\beta$ in source B. Note that the [O\,{\sc iii}] and the narrow component of H$\beta$ are constrained to have the same profile.  
When we include additional outflow components for the [O\,{\sc iii}] doublet in the two sources (orange curves in the right column of figure \ref{fig:2337}  panel (c)), the fit is significantly improved. 
Based on [O\,{\sc iii}], we measured a redshift of 0.7089 for source A and 0.7083 for source B. The velocity offset of the narrow lines between two sources is $102 \pm 7\,\mathrm{km\ s^{-1}}$ . This, together with the different line ratios, cause us to classify the target as a dual quasar system.\par
\par
J2337 is the only target for which we have to fit an iron template to reconstruct the spectra. The relative strength of the iron emission in source B is higher than that in source A, while the equivalent width (EW) of [O\,{\sc iii}] core component (magenta components) and FWHM of H$\beta$ are narrower in source B. This follows the trend expected by  Eigenvector 1 (e.g., \citealt{shen2014diversity}).

\subsubsection{SDSS J123821.66+010518.6} \label{subsubsec:J1238}
\begin{figure*}[htp]
\centering
\includegraphics[width=1\textwidth]{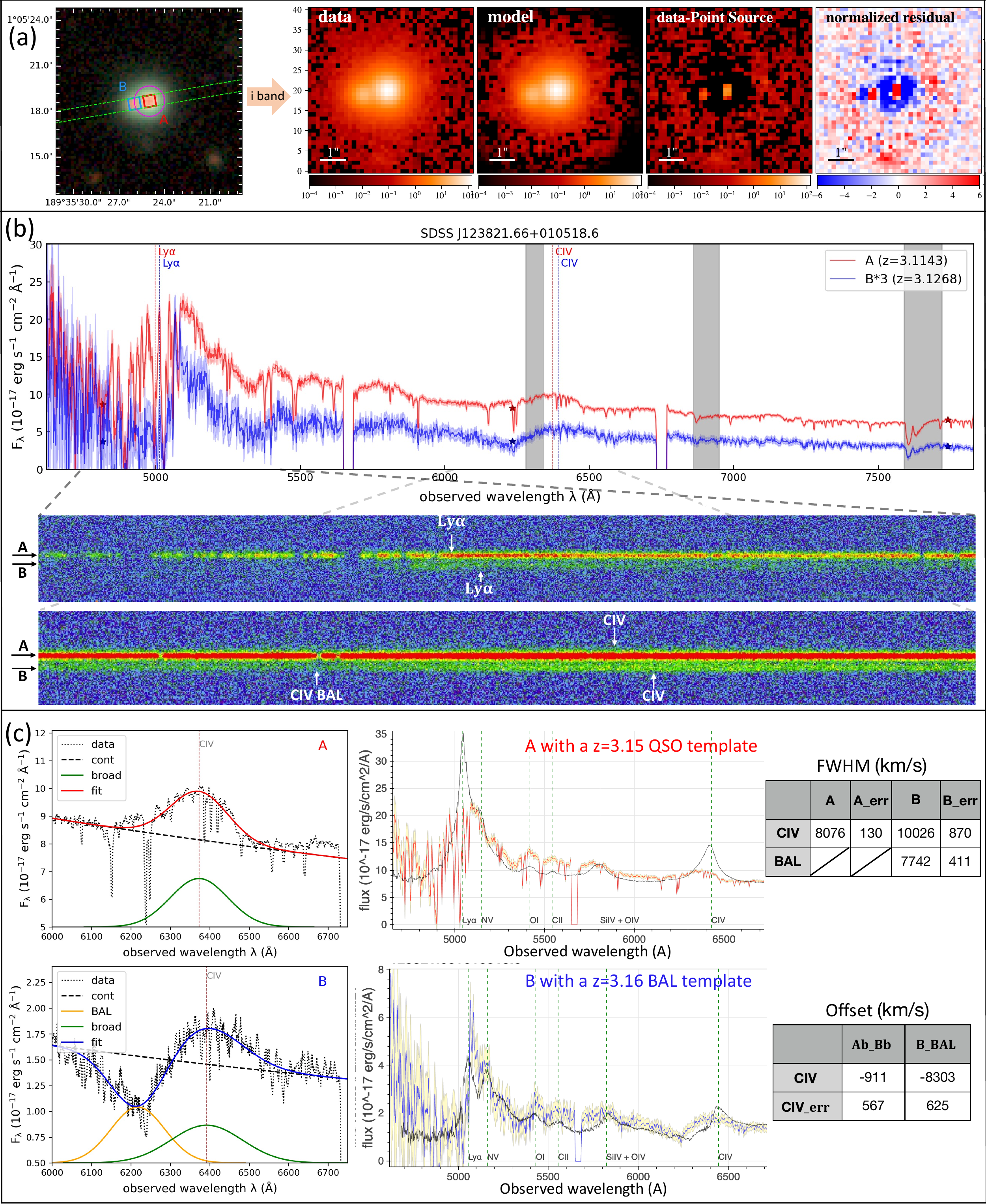}
\caption{A kaleidoscope of SDSS J1238+0105. Panel (a): HSC imaging overlaid with spectroscopic setups and decomposition results on $i$ band image. Panel (b): 1D and 2D spectrum, the 1D spectrum of each source is rescaled to match with the PSF magnitudes labeled with star symbols respectively. The grey shadows indicate the telluric absorption bands. Panel (c): Supplementary information. Both the line fitting results and template matching results are shown. The tables show the offsets and FWHMs of the C\,{\sc iv} emission line and BAL.
\label{fig:1238}}
\end{figure*}
The SDSS BOSS program reports this target as a broad-line quasar at $z = 3.1327$. Both sources are red in $g-r$ color (Figure \ref{fig:1238}a) and lie close to the stellar locus (Figure \ref{fig:distance}). The primary source is two magnitudes brighter than the secondary. The host galaxies are undetected in this system in our image decomposition, which is unsurprising given the redshift of the pair. 
\par
The GMOS spectrum of this target was taken with the B600+CG455 grism. Ly$\alpha$ and C\,{\sc iv}$\lambda$1549 were both observed (Figure \ref{fig:1238}b), and the latter was used to measure the redshifts of the sources. However, we found the redshifts measured via line fitting is not perfectly consistent with template matching. In Figure \ref{fig:1238}c we show that the redshifts preferred by the SDSS templates\footnote{\url{http://classic.sdss.org/dr5/algorithms/spectemplates/}} are actually 3.15 for source A (template No.30) and 3.16 for source B (template No.32). We suggest the asymmetry of the C\,{\sc iv}$\lambda$1549 line is the reason for this difference. 
\par
Furthermore, we found that the companion source has the  `P Cygni' profile typical of BAL quasars, while the primary source does not. We fit the profile of the companion with two Gaussian functions (Figure \ref{fig:1238}c left figures), and find that the BAL component is blueshifted from the emission line by $8303 \pm 625\,\mathrm{km\ s^{-1}}$ with a FWHM of $7742 \pm 411\,\mathrm{km\ s^{-1}}$, compared to  $10026 \pm 870\,\mathrm{km\ s^{-1}}$ for the emission-line component. According to \cite{weymann1991comparisons}, a quasar with contiguous absorption extended over more than 2000 km/s and with a blueward velocity of at least 5000 km/s can be considered as a BAL quasar. Therefore, we classify this as a BAL quasar. 
\cite{hamann1993geometry} studied line profiles of 40 BAL quasars, and found that most BAL quasars have stronger absorption than the emission, which suggests that the BAL regions do not completely envelope the continuum source. For J1238B, the equivalent width of the C\,{\sc iv}$\lambda$1549 emission line is 55 \text{\AA}, smaller than 60 \text{\AA} of the BAL. Also, there are multiple narrow absorption lines in the brighter source, while they are absent in the fainter source, which tells us that the absorption clouds are in front of only the brighter source only.

This is the highest redshift quasar pair at close projected separation (7.67 kpc) yet found. 
Fortuitously, the BAL nature of the companion firmly rules out the possibility of it being a lensed system thus mitigating such complications \citep{shen2021hidden}.

\section{Discussion} \label{sec:Discussion}
In this section we discuss the physical properties of our dual quasar candidates and compare those properties to single quasars, normal galaxies, and simulation results. Our aim is to further understand the role that dual quasars play in the scenario of galaxy mergers, and examine how mergers impact the SMBH--host galaxy correlation.
\subsection{Broad-Line Region and black hole mass estimates}
For type-1 quasar pairs, we estimate the BH masses for each source using the viral method \citep{vestergaard2006determining}: 

\begin{equation}
M_{\mathrm{BH}}=10^{\alpha} L^{\beta} \left(\frac{\mathrm{FWHM}}{1000~\mathrm{km}~ \mathrm{s}^{-1}}\right)^{\gamma} M_{\odot}
\label{eq:viral}
\end{equation}

\noindent where the FWHM is the full-width half-maximum of the specific broad emission lines used for the calculation, $L$ is the corresponding monochromatic luminosity, and $\alpha$, $\beta$, $\gamma$ are the indices calibrated from reverberation mapping methods. Our measurements are based on C\,{\sc iv}, Mg\,{\sc ii}, H$\beta$, and H$\alpha$. The parameters used for each are listed in Table \ref{tab:viral}. For each observed line, the monochromatic luminosity is measured from the power-law fit to the continuum, and the FWHM is measured from Gaussian fit described in Section \ref{sec:Fitting}. We further note that the normalization of the spectra are set by the HSC photometry after running our 2D decomposition. Thus the quasar luminosity is free of host galaxy contamination.

\begin{deluxetable}{ccccc}
\tablecaption{Parameters of equations \ref{eq:viral}}
\tablewidth{1\textwidth}
\tablenum{4}
\tablehead{
\colhead{line} & \colhead{L} & \colhead{$\alpha$} & \colhead{$\beta$} & \colhead{$\gamma$}
}
\startdata
C\,{\sc iv} & $L_{1350,44}$ & 6.66 & 0.5 & 2\\
Mg\,{\sc ii} & $L_{3000,44}$ & 6.74 & 0.62 & 2\\
H$\beta$ &  $L_{5100,44}$ & 6.91 & 0.5 & 2\\
H$\alpha$ &  $L_{H\alpha,42}$ & 6.71 & 0.48 & 2.12\\
\enddata
\tablecomments{Parameters used for black hole mass estimation in Equation \ref{eq:viral}, referring to  \cite{vestergaard2006determining, schulze2017near, shen2011catalog}. For the second column, $L_{1350,44}$ stands for the monochromatic luminosity at 1350 \text{\AA} in units of $10^{44}\,\mathrm{erg}\,\mathrm{s}^{-1}$, and similar for the others.
\label{tab:viral}}
\end{deluxetable}

\begin{figure}[htp]
\centering
\includegraphics[width=0.5\textwidth]{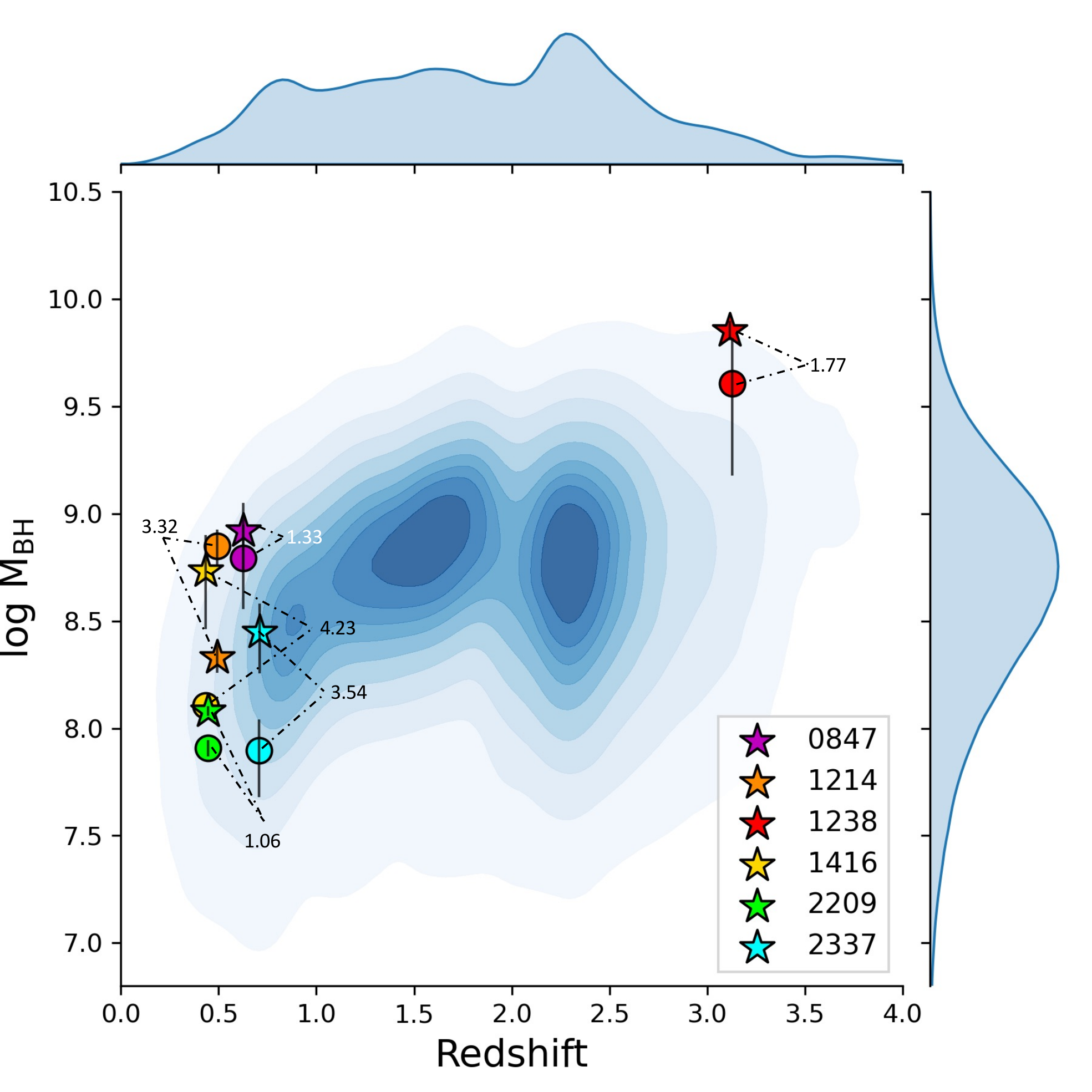}
\caption{Quasar pairs in BH mass vs redshift space of the HSC-SDSS dual quasars (colored stars). When there are multiple estimations of BH mass from different emission lines for a single object, we show the median value. The two objects of a pair are indicated with the same color, star symbols for the primary source and the circular symbols for the companions, connected with dot-dashed lines showing the mass ratio of the pair. For comparison, quasars from the whole DR14 quasar catalog are shown with contours determined via KDE and projected to the marginal curves.}
\label{fig:BH_mass}
\end{figure}

\begin{figure*}[htp]
\centering
\includegraphics[width=1\textwidth]{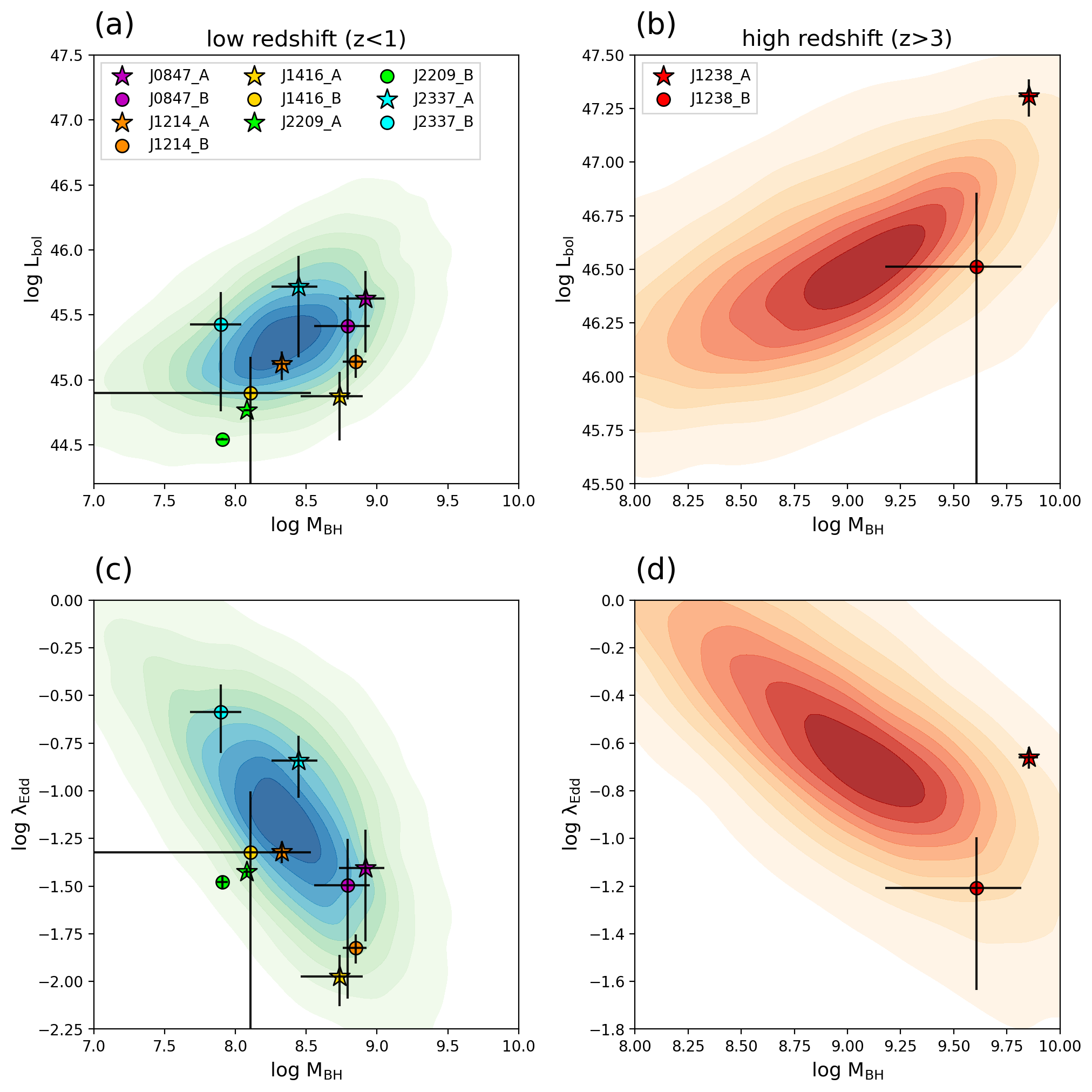}
\caption{Bolometric luminosity and Eddington ratio as a function of black hole mass on log scales. The comparison sample is again the whole SDSS DR14 catalog,  This time we separate the comparison into redshift bins: $z<1$ and $z>3$, recognizing the different population of quasars at low and high redshifts. We do not include quasars with $1 < z < 3$ since we have no quasar pair confirmed in that redshift range. Our quasar pairs are plotted using the same symbols as in Figure \ref{fig:BH_mass}. Panels (a) and (b) show the distribution in $L_{\mathrm{bol}}-M_\mathrm{BH}$ space, while panels (c) and (d) are plotted in $\lambda_{\mathrm{edd}}-M_{\mathrm{BH}}$ space.}
\label{fig:bol_edd}
\end{figure*}

In Figure \ref{fig:BH_mass}, we plot the black hole mass for the objects in our sample as a function of the spectroscopic redshift, and label both quasars in each dual quasar system with the same colored symbol. We also provide the value of the mass ratio of the two, connected by the dashed lines in the figure. The measurement errors on the black hole masses are propagated from the uncertainties on the parameters as output from the line-fitting routine. Also, the virial estimators have an intrinsic scatter of 0.3 dex. The errors on the redshift measurements are negligible. For comparison, we use the values for the full SDSS quasar sample as provided by \cite{rakshit2020spectral} as indicated by the background sources with contours indicative of the number density of quasars. The comparison sample includes 525,103 quasars from SDSS DR14 with black hole mass measurements. The full quasar sample covers a redshift range from 0 to 4.0 and black hole masses between $10^7$ to $10^{10}$ $M_{\odot}$. Our dual quasars have masses comparable to single quasars and have black hole mass ratios within a factor of five (Figure \ref{fig:BH_mass}).
\par
Figure \ref{fig:bol_edd} displays the bolometric luminosity and Eddington ratio of our dual quasars as a function of black hole mass. The log bolometric luminosity ranges from 44.2 to 47.5, while the log Eddington ratio ranges from $-2.2$ to 0. The bolometric luminosity $L_{\mathrm{bol}}$ in the upper panel of Figure \ref{fig:bol_edd} is calculated using monochromatic luminosity and corresponding bolometric correction factors provided by \cite{richards2006spectral}: BC5100 = 9.26, BC3000 = 5.15, and BC1350 = 3.81. The Eddington luminosity $L_{\mathrm{Edd}}$ is estimated from the BH mass as:
\begin{equation}
L_{\mathrm{Edd}} \cong 1.3 \times 10^{38}\left(M_{\mathrm{BH}} / M_{\odot}\right) \mathrm{erg}\,\mathrm{s}^{-1}
\end{equation}
The Eddington ratio is then given by $L_{\mathrm{bol}}/L_{\mathrm{Edd}}$. We separate our quasar pairs into low-redshift ($z<1$) and high-redshift ($z>3$) pairs and compare with the corresponding SDSS population in Figure \ref{fig:bol_edd}. Panels (a) and (c) show the low-redshift cases, while (b) and (d) show the high-redshift objects. 

\par
We find that the bolometric luminosity and Eddington ratio of both objects in the low-redshift quasar pairs 
do not have significant difference from single quasars at the same BH mass. This agrees with the results of \cite{stemo2020catalog}. Figure \ref{fig:z_success} shows that we do not bias to equally luminous pairs in follow-up observation. However the selection algorithm itself could be biased to candidates with stronger point source features, thus more likely to have similar properties with the single quasars when successfully confirmed. 
For the high-redshift pair (SDSS J1238+0105), the main source is luminous with a high Eddington ratio, 
while the fainter source has large uncertainties in the measurements, making it harder to draw a strong conclusion.


To highlight, we have so far not found any quasar pairs in the redshift range between 1 to 3 even with suitable candidates as shown in Figure~\ref{fig:z_success}. In this redshift range, our spectra are only able to cover one emission line, either C\,{\sc iv} $\lambda$1549 or Mg\,{\sc ii} $\lambda$2798, hence the identification of a weak emission line is difficult,  especially in case of type 2 quasars.


\subsection{Black Hole Growth Relative to Galaxy Growth in Mergers with Dual Quasars} \label{subsec:region}

Our decomposition algorithm returns five-band photometric magnitudes of the host galaxies of our dual quasar systems. Based on this photometry, we carried out a stellar mass estimation for systems at $z<1$ using CIGALE (Code Investigating GALaxy Emission \cite{boquien2019cigale}). \cite{li2021sizes} used CIGALE models for the study of quasar hosts with HSC. The computation of stellar mass includes an assumed star formation history (SFH), nebular emission, attenuation, dust emission (not included in our cases due to the lack of IR data), and AGN emission (not included in our cases because we have already subtracted the point sources) to generate a model SED based on a fit to the HSC photometry.

\begin{figure}[htp]
\centering
\includegraphics[width=0.5\textwidth]{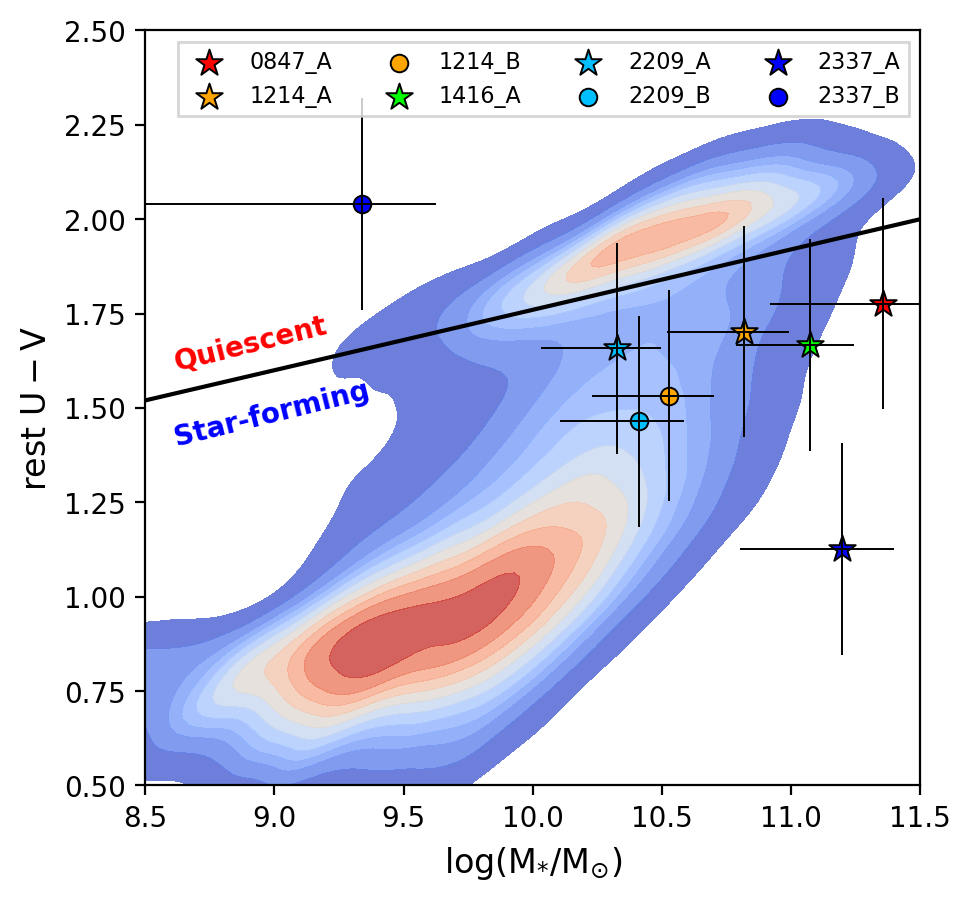}
\caption{Host galaxy properties of our dual quasars in rest-frame $U - V$ vs stellar mass diagram. The fitting model is either 2 PSF + 1 S\'ersic profile (for objects with a common envelope) or 2 PSF + 2 S\'ersic profile. The star symbols correspond to the main quasars' hosts and the circular symbols with the same color correspond to the companions' hosts. For objects modeled with one galaxy component, only the star symbols are labeled. The density map shows a sample of 709,006 normal galaxies over the redshift range 0.2 to 1 with $10^{7} < M_* < 10^{13}$ selected from the second HSC public data release (Kawinwanichakij et al. in prep). The boundary line is $U − V = 0.16 \times \mathrm{log}\ M_{*} + 0.16$, derived with the Support Vector Machine algorithm, separate the galaxies into star-forming and quiescent \citep{li2021sizes}.
\label{fig:color_Ms}
}
\end{figure}

In Figure \ref{fig:color_Ms}, we show the rest-frame $U - V$ color of the host galaxies of our dual quasars as a function of the stellar mass. Recall that we modeled either one or two galaxy components for the pairs. In Figure \ref{fig:color_Ms}, the star symbols represent the hosts of the main quasars, while the circular markers of the same color represent the companions if two separate galaxy components are identified. The $U$ and $V$ band magnitudes are measured from the best fit SED model. A typical color uncertainty of 0.28 mag is applied to all the sources, and the uncertainties of the stellar masses are given by CIGALE. For comparison, we provide a sample set of 709,006 normal galaxies over the redshift range 0.2 to 1 with $10^{7} < M_* < 10^{13}$ (Kawinwanichakij et al. in prep). We find that the host galaxies of our quasar pairs are mostly massive, but are bluer than red sequence galaxies. Most of them are located below the boundary line $U − V = 0.16 \times \mathrm{log}\ M_{*} + 0.16$, which separates the star-forming and quiescent galaxies. \cite{capelo2015growth} suggested in their simulation that the interaction of galaxies may trigger star formation in the merger stage (see their Figure 1 and 2). This also agrees with the observational findings of host galaxies of single quasars \citep{matsuoka2015sloan,ishino2020subaru,li2021sizes}.

\subsection{BH-host mass correlation from the Horizon-AGN simulation and our observations}
\begin{figure*}[htb]
\centering
\includegraphics[width=1\textwidth]{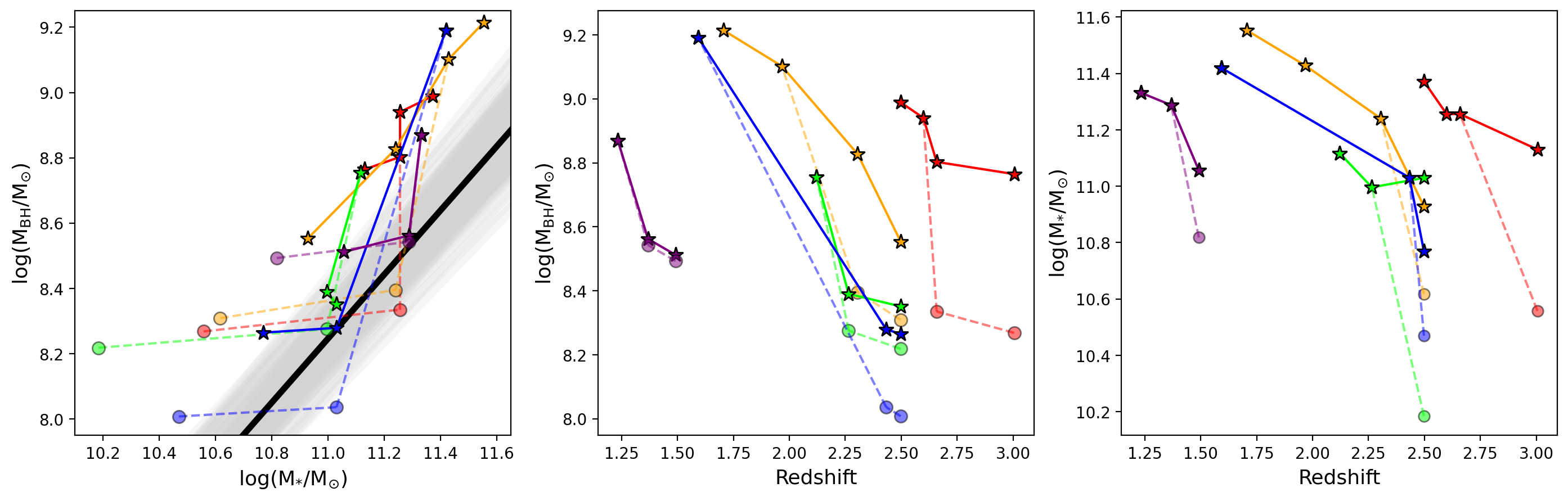}
\caption{Evolutionary trajectories of five AGN pairs obtained from the Horizon-AGN simulation. Left panel: evolution of $M_{\mathrm{BH}}-M_{*}$ during four stages of the merger. The colors indicate the different AGN pairs. The stars mark the more massive source of the pairs, connected by solid lines to show their evolutionary trajectories. Dots show the less massive companions, evolving with dashed lines. The BHs form a unresolved binary at the third stage, and are indicated by a single star thereafter. The thick black line is the best-fit model of observational data from \cite{haring2004black}, while 1$\sigma$ confidence interval is indicated by the grey region. Middle panel: evolution of BH mass with redshift. Right panel: evolution of stellar mass with redshift.
\label{fig:horizon}
}
\end{figure*}

The correlation between the BH mass and mass of the host galaxy has been quantified by many studies, but the reason for this connection remains uncertain. 
\cite{silk1998quasars, fabian1999obscured, king2003black}, among others suggest that mergers work as a global process to build this correlation. If this is true, we may expect to see that systems prior to merging are offset from the local correlation, with the merger bringing them onto the correlation. However, it is difficult to trace the full life cycle of a merger with observational snapshots. Therefore, we will first test this hypothesis from the view of the Horizon-AGN simulation.
\par
Horizon-AGN is a cosmological hydrodynamical simulation with the size of the box $L_{\mathrm{box}} = 100 h^{−1}$ Mpc. \cite{dubois2014dancing} measured the properties of their mock galaxies and studied the correlation between the spin directions and the large scale cosmic web structure.  \cite{volonteri2016cosmic} included BHs and AGN in their simulation.  BHs are created in gas and stellar over-density regions that exceed the threshold of star formation, $n_0=0.1\,\mathrm{H\,cm^{-3}}$, the seed mass is assumed to be $10^5~\mathrm{M_{\odot}}$. The accretion onto BHs follows the Bondi-Hoyle-Lyttleton (BHL) accretion rate: 
\begin{equation}
\dot{M}_{\mathrm{acc}}=\frac{G^{2} M_{\mathrm{BH}}^{2} \rho_{\mathrm{gas}}}{\left(c_{s}^{2}+v_{\mathrm{rel}}^{2}\right)^{3 / 2}}
\end{equation}
where $M_{\mathrm{BH}}$ is the BH mass, $\rho_{\mathrm{gas}}$ is the average gas density, $c_s$ is the average sound speed, $v_{\mathrm{rel}}$ is the average gas velocity relative to the BH velocity.\par
A BH is assigned to a halo when it is located within $0.1 \times R_{\mathrm{vir}}$ of the center (central BH) or within $R_{\mathrm{vir}}$ (off-center BH). They found that a fraction of galaxies can host both a central BH and an off-center BH, thus forming a binary system. When both BH have $L_{\mathrm{bol}}>10^{43} \mathrm{erg\ s^{-1}}$, they form an AGN pair. 
\par
We successfully traced the evolution of five such AGN pairs from Horizon-AGN simulation with 
primary quasar luminosity $> 10^{45.3}$, and the secondary $> 10^{44.3}$ (M. Volonteri, private communication). The real space separations of these sources are between 5 and 30 kpc. These criteria are the same as those we used to select our observational candidates. This left us with five pairs for which we are able to trace the full merger sequence (Figure \ref{fig:horizon}). Star symbols represent the primary BH (central BH), and the dots represent the secondary BHs (off-centre BH); each pair is marked with a separate color.   The BH mass and galaxy stellar mass are recorded in four stages in the merger scenario: (1) initial stage when both sources are labeled as AGN and are well separated. (2) The host galaxies merge first. (3) The BHs form a close SMBH binary separated by 1 kpc or less, where they become unresolved at this stage, thus the $M_{\mathrm{BH}}$ is the total mass of the pair, and (4) They coalesce into one single source. Figure \ref{fig:horizon} shows the evolution trajectories of these five AGN pairs. In some cases, the final coalescence stage is very close to the SMBH binary stage, thus only three stages are visible. The thick black line is a fit to the $M_{\mathrm{BH}}-M_*$ correlation based on the observational data set of local AGNs (from \cite{haring2004black}):
\begin{equation}
\log \left(\mathrm{M}_{\mathrm{BH}} / 10^{7} \mathrm{M}_{\odot}\right)=0.27+0.98 \log \left(\mathrm{M_*} / 10^{10} \mathrm{M}_{\odot}\right)
\end{equation}
The 1$\sigma$ confidence interval is indicated in grey. 

\begin{figure}[htb]
\centering
\includegraphics[width=0.5\textwidth]{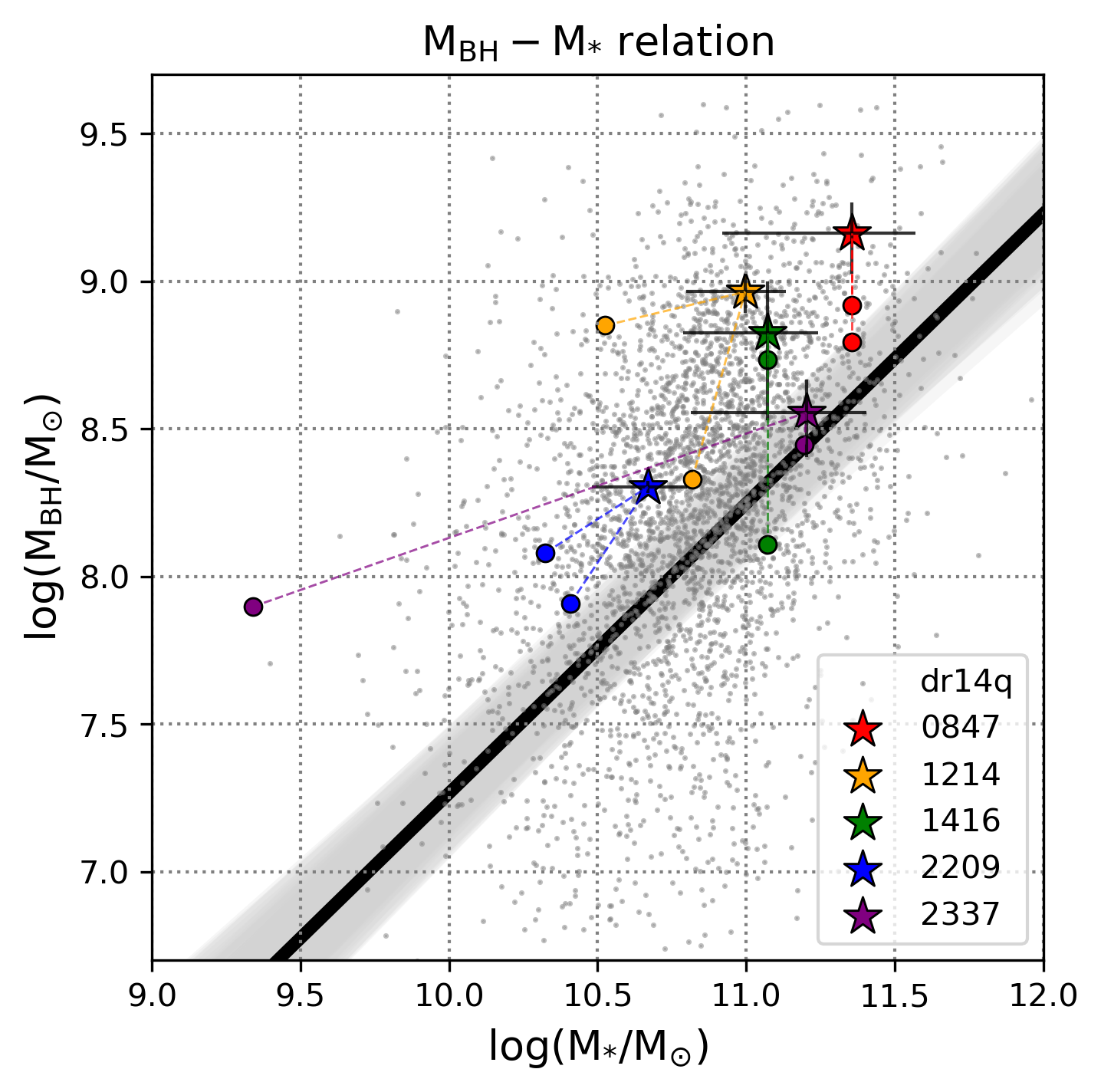}
\caption{Comparison of our dual quasar samples with single quasars and the local BH-host correlation. The thick black line and grey shading are the same as that in figure \ref{fig:horizon}. 3,315 single quasars are selected from SDSS DR14 (details in \cite{li2021sizes}), with BH mass taken from \cite{rakshit2020spectral}, and host mass calculated in the same way with decomposition and SED fitting. Our dual quasars are plotted with the dots representing the two BHs separately, and the star symbols are the sum of the dots, mimicking the final state of the merger.}
\label{fig:MBH_Ms}
\end{figure}
\par
To compare our data with the evolutionary traces in Horizon-AGN, we plot black hole mass and stellar mass based on the image decomposition of our dual quasars in Figure \ref{fig:MBH_Ms}. In total, we are able to measure BH masses and stellar masses for five quasar pairs: J0847, J1214, J1416, J2209, and J2337. Here our quasar pairs are plotted as colored points, and circles with the same color for the two sources in a pair. Unlike the simulation, we can only capture one phase in the merger process. As a highly simplified exercise, we sum the values of the two components in each pair as indicated by the star symbols, mimicking their possible final states after coalescence. Note that in two cases: J0847 and J1416, their host galaxies cannot be separated, the model prefers to fit the image with one Sérsic profile, thus we have only one measurement for the stellar mass. The grey dots are 3,315 single quasars at $0.2 < z < 1.0$ selected from SDSS DR14 (details in \cite{li2021sizes} for comparison. We found that our dual quasars mostly have an over-massive BHs compared to the local correlation, and this excess keeps after the merger. This is partly due to the bias of the SDSS survey that only the most luminous and massive quasars are detected.
\par
We found similar tendencies in the left panel of Figure \ref{fig:horizon} and in our results (Figure \ref{fig:MBH_Ms}). Either before or after the merger, most of the dual AGN in the simulation are located above the local correlation. This indicates that the BHs have already grown to a massive state before the merger. This agrees with the distribution of our data in Figure \ref{fig:BH_mass}, showing the dual quasars possess  BHs with masses comparable to single qusars. In fact, the mergers do not bring them any closer to the correlation; rather, the merger products lie even further above the correlation.  
we trace the evolution of the Horizon AGNs with redshift in Figure \ref{fig:horizon} (middle and right panels). We find that the mergers do lead to a boost in the BH mass (see where the dashed lines and solid lines merge), especially the blue case. However, the gain in stellar mass is not as dramatic. In one case (the green one), we even see a {\it decrease} in stellar mass during the merger. We suggest that during the merger, the systems lost matter due to the interaction, stellar feedback caused by starburst or AGN-driven effects, while the central black holes of course do not lose mass. This is 
why we see the mergers bring the AGN pairs above the local correlation. However, biases may still be present in these results: 
the less massive (i.e., fainter) BHs could be missed by our selection algorithm, which prefers point-like sources. When the central source is not bright enough to outshine its host, we may fail in identifying such candidates. Also, with our spectroscopic observations, we may fail to identify a low-mass BH when the spectrum is too faint to recognize the emission lines.
\par
A larger observational sample is required to make more definitive statements about the relative evolution of black holes and their hosts in galaxy mergers with dual black hole growth. In a future study, we will expand our sample size both from candidate selection and additional follow-up spectroscopic observations. The HSC/SSP is continuing to release new data thus we plan to update our parent samples from PDR2 to S20A (DR3). Together with a more refined selection algorithm, we expect to grow our parent sample by a fator of 2. We have also achieved new data from the subsequent Gemini/GMOS queue and the 3.58m ESO New Technology Telescope (NTT). We will report our findings in follow-up works.

\section{CONCLUSION} \label{sec:Conclusion}
There is much interest in understanding the frequency of dual quasar activity in our Universe for understanding the growth of both supermassive black holes and their host galaxies. In particular, dual quasars with separations closer than 30 kpc play an important role in  unveiling possible mechanisms to form and trigger quasars, and explaining the gravitational wave background in the nanohertz (nHz) band. Following our previous work \cite{silverman2020dual}, this study enlarges the sample size of the dual quasars in this effort, and provides the details of our data analysis procedures. We first summarize the observational findings of this work as follows:

\begin{itemize}
    \item We utilized a method to systematically search for closely separated ($0^{\prime\prime}.6-4^{\prime\prime}$) dual quasar candidates based on Subaru/HSC-SSP imaging data, and carried out follow up spectroscopic observations to confirm their dual nature and to measure the fraction of the candidates that are in fact dual (Section \ref{subsec:Rate}).
    \item We implemented a method to separate quasars from stars in the color-color diagram based on the distance to the stellar locus. In a specific region of this diagram, we found 7 dual quasars (including physically-associated and projected quasar pairs) out of 9 candidates (Section \ref{sec:Results}).
    \item We acquired spectroscopic data for 32 candidates, among which we discovered 6 physically-associated quasar pairs (Section \ref{sec:Results}).
    \item We compared the colors of our dual quasars to those of SDSS quasars, and found that these dual systems tend to be redder, which could be caused by dust extinction from the merger itself (Figure \ref{fig:SDSS_color}).
\item We identified the highest redshift ($z=3.1$) dual quasar with secure spectroscopic confirmation and having a projected physical separation less than 10 kpc.

\end{itemize}

Based on these observational results, we carefully estimated the physical properties of our dual quasar samples and compared with recent simulation results, giving a first view on the impact of mergers on BH-host correlation. We summarize our scientific findings of our dual quasar as follows:
\begin{itemize}
    \item We estimated the virial BH masses of our dual quasars 
    and found that they have already reached values of the overall SDSS BH mass population. We found all the pairs have black hole masses with ratios of 5:1 or less (Figure \ref{fig:BH_mass}).
    \item We estimated the bolometric luminosities and Eddington ratios of our dual quasars using monochromatic luminosities and standard bolometric correction factors. We found these properties of the pairs are also comparable to the typical single quasars in the SDSS catalog, which might be due to the bias of our selection algorithm, which picks out the most point-like sources (Figure \ref{fig:bol_edd}).
    \item Taking advantages of high quality imaging data from Subaru/HSC, we estimated the stellar mass of the host galaxies of our dual quasars using CIGALE, which we compared to normal galaxies. We found most of them are massive galaxies $~10^{11}M_{\odot}$ that bluer than the red sequence, which could be due to an increasing star forming activity (Figure \ref{fig:color_Ms}).
    \item We investigated their position relative to the $M_{\mathrm{BH}}-M_{*}$ correlation of our quasar pairs and compared to five quasar pairs selected from the Horizon-AGN simulation. We found that they agree with each other in that both reside above the local relationship, either before or after final coalescence. This indicates that the $M_{\mathrm{BH}}-M_*$ correlation is possibly not established via mergers as some studies suggest (Figure \ref{fig:MBH_Ms}).
\end{itemize}

\begin{acknowledgments}
The Hyper Suprime-Cam (HSC) collaboration includes the astronomical communities of Japan and Taiwan, and Princeton University.  The HSC instrumentation and software were developed by the National Astronomical Observatory of Japan (NAOJ), the Kavli Institute for the Physics and Mathematics of the Universe (Kavli IPMU), the University of Tokyo, the High Energy Accelerator Research Organization (KEK), the Academia Sinica Institute for Astronomy and Astrophysics in Taiwan (ASIAA), and Princeton University.  Funding was contributed by the FIRST program from the Japanese Cabinet Office, the Ministry of Education, Culture, Sports, Science and Technology (MEXT), the Japan Society for the Promotion of Science (JSPS), Japan Science and Technology Agency  (JST), the Toray Science  Foundation, NAOJ, Kavli IPMU, KEK, ASIAA, and Princeton University.
\par
This paper makes use of software developed for the Large Synoptic Survey Telescope. We thank the LSST Project for making their code available as free software at  http://dm.lsst.org
\par
This paper is based [in part] on data collected at the Subaru Telescope and retrieved from the HSC data archive system, which is operated by Subaru Telescope and Astronomy Data Center (ADC) at NAOJ. Data analysis was in part carried out with the cooperation of Center for Computational Astrophysics (CfCA), NAOJ.
\par
The Pan-STARRS1 Surveys (PS1) and the PS1 public science archive have been made possible through contributions by the Institute for Astronomy, the University of Hawaii, the Pan-STARRS Project Office, the Max Planck Society and its participating institutes, the Max Planck Institute for Astronomy, Heidelberg, and the Max Planck Institute for Extraterrestrial Physics, Garching, The Johns Hopkins University, Durham University, the University of Edinburgh, the Queen’s University Belfast, the Harvard-Smithsonian Center for Astrophysics, the Las Cumbres Observatory Global Telescope Network Incorporated, the National Central University of Taiwan, the Space Telescope Science Institute, the National Aeronautics and Space Administration under grant No. NNX08AR22G issued through the Planetary Science Division of the NASA Science Mission Directorate, the National Science Foundation grant No. AST-1238877, the University of Maryland, Eotvos Lorand University (ELTE), the Los Alamos National Laboratory, and the Gordon and Betty Moore Foundation.
\par
The spectroscopic data are taken in Subaru/FOCAS program S19B-079 in remote mode, Gemini/GMOS program GN-2019B-Q-128 in queue mode, we would like to thank the support astronomers for observation. Tang wants to thank T. Hattori for his help on FOCAS data reduction and the Pypeit team for their help on GMOS data reduction. J.D.S. is supported by the JSPS KAKENHI Grant Number JP18H01251, and the World Premier International Research Center Initiative (WPI Initiative), MEXT, Japan.
\end{acknowledgments}

\bibliography{myarticle}{}
\bibliographystyle{aasjournal}

\appendix

\section{Calibration of stellar locus from SDSS to HSC} \label{sec:Calibration}
\begin{figure}[htp]
\centering
\includegraphics[width=120mm]{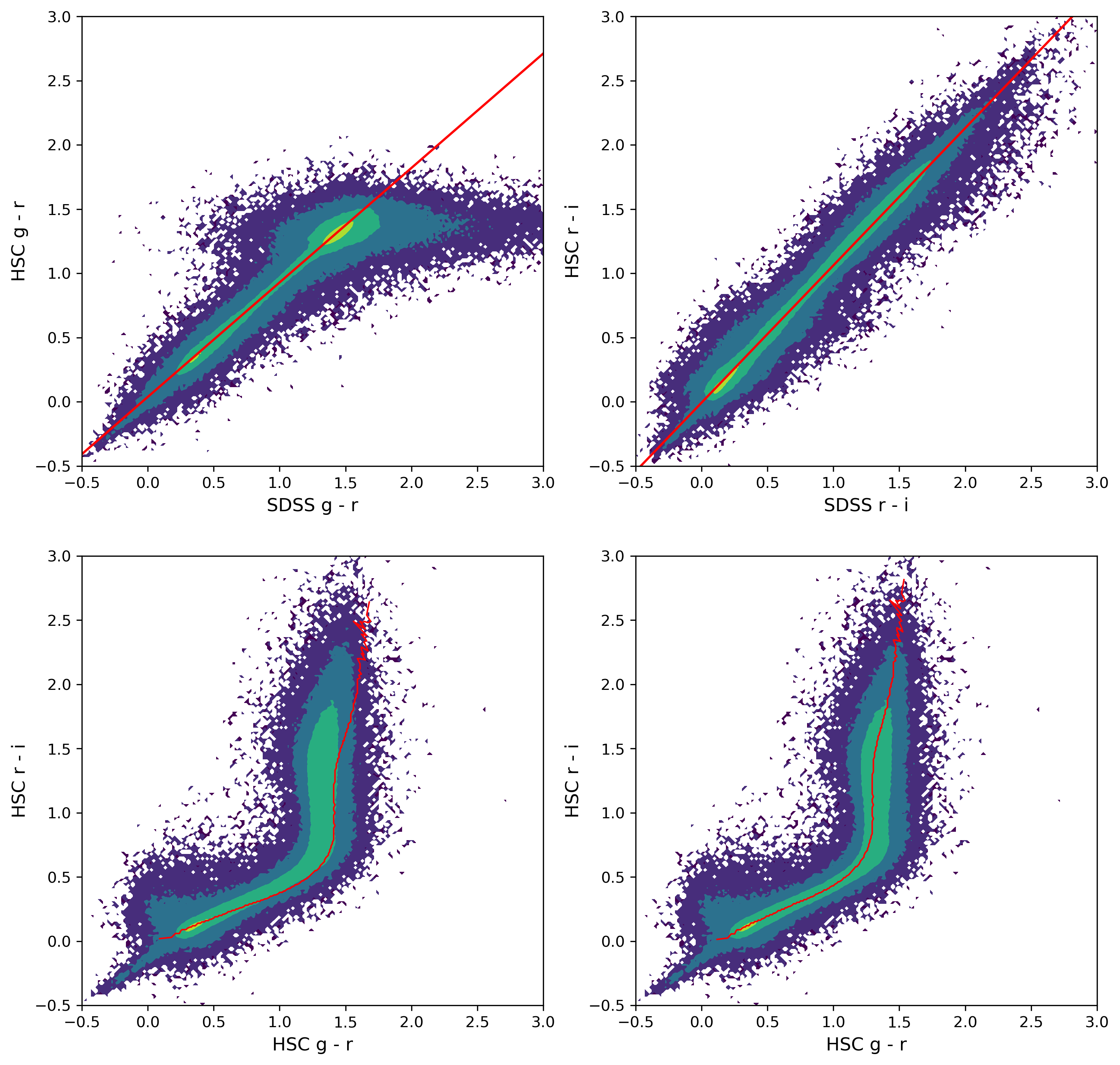}
\caption{Calibration of the stellar locus to the HSC photometric system. Upper panels: linear regression applied to HSC and SDSS colors, red lines are the regression results, contours are a set of stars queried from Stripe 82 data. Lower panels: stellar locus of \cite{covey2007stellar} before and after correction to HSC, contours are the same set of data as above.
\label{fig:remake}}
\end{figure}

Through our 2D image analysis, we acquired five band PSF magnitudes of both components in each potential dual quasar system. We further identify high priority targets based on both its color and its distance from the stellar locus, which represent the typical values of the colors of stars \citep{covey2007stellar}. However, the photometric bands of SDSS and HSC have different sensitivity curves, which call for a correction on the stellar locus from SDSS color to HSC color. The SDSS stellar locus is plotted as red curve in Figure \ref{fig:remake} bottom left panel, while the contours are made by a set of 619,634 stars queried from stripe 82 catalog \citep{annis2014sloan}. The stars are matched to HSC catalog to get the HSC colors. As expected, the SDSS stellar locus has a significant shift from the data set based on HSC photometry. The correction are made assuming a linear relationship between SDSS and HSC color, i.e.
\begin{equation}
    (g-r)_{\mathrm{HSC}} = k1 * (g-r)_{\mathrm{SDSS}} + b1
\end{equation}
\begin{equation}
    (r-i)_{\mathrm{HSC}} = k2 * (r-i)_{\mathrm{SDSS}} + b2
\end{equation}
The upper panels of figure \ref{fig:remake} show the results of linear regression indicating k1 = 0.90093, b1 = 0.02389, k2 = 1.11105, b2 = -0.03081. The lower right panel shows the fitted stellar locus, which now matches HSC data better. We use this corrected stellar locus to study the color dependence of our candidates in Section \ref{sec:Results}.\par

\section{By-products} \label{sec:by_products}

\begin{figure*}[htb]
\centering
\includegraphics[width=1\textwidth]{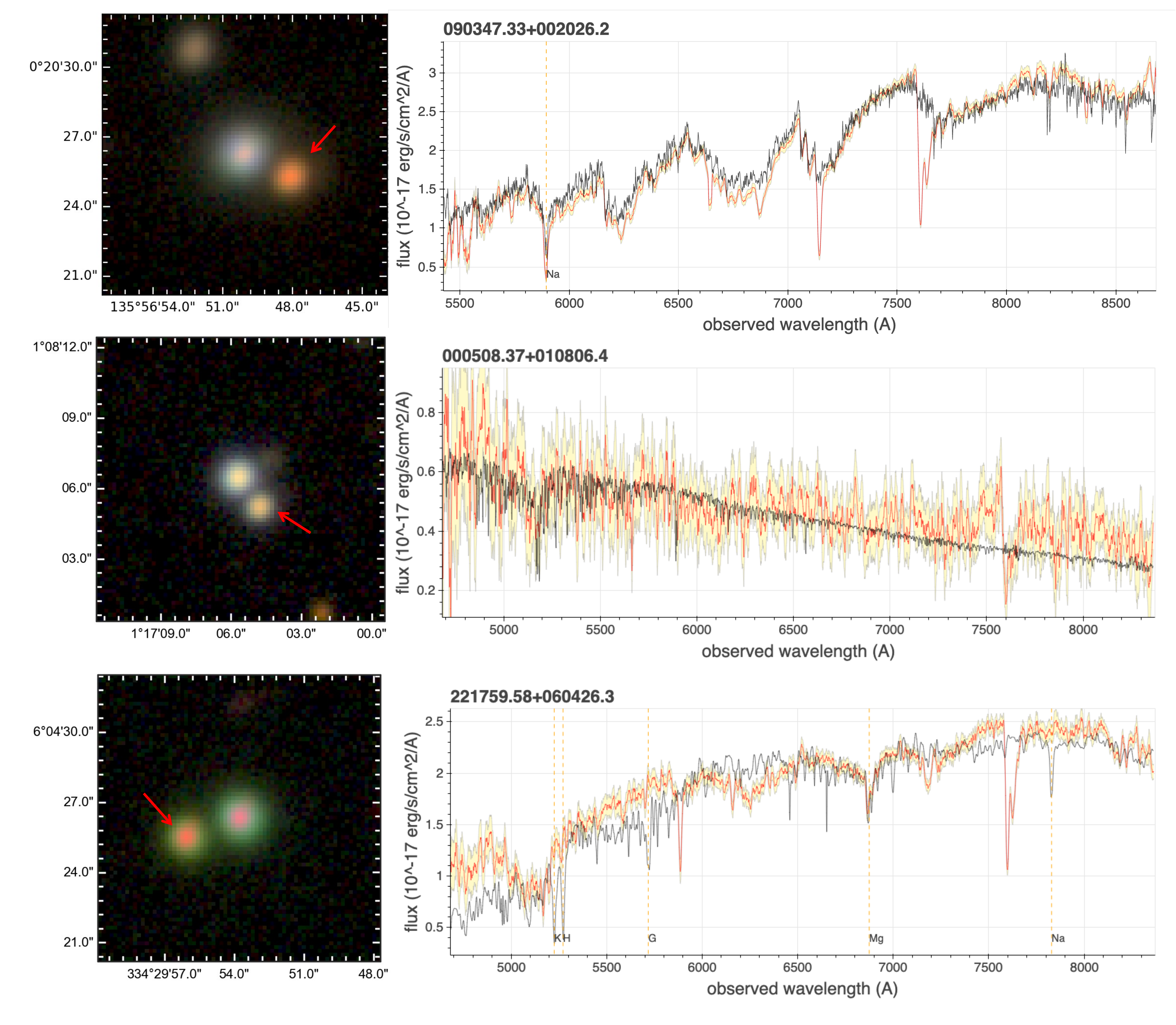}
\caption{Examples of template matching, together with the HSC images (colored according to \cite{lupton2004preparing}). Red curves are the data of the companion source (pointed by red arrows in HSC images), shadows show the 1$\sigma$ region. Black curves are SDSS templates, orange vertical dashed lines label the position of some main absorption lines. Top: M1 star spectrum. Middle: K-type star spectrum. Bottom: Early-type galaxy spectrum shifted to $z=0.328$. 
\label{fig:bokeh}}
\end{figure*}

Here, we present by-products from this work, i.e., candidates that are not confirmed as physically associated quasars. Compared to quasars, whose features are easy to identify even by eye, the spectra of stars and galaxies can be challenging to classify in some cases, especially when they are featureless. To classify these targets, we use the cross-correlation templates (\cite{bolton2012spectral}, also at \url{http://classic.sdss.org/dr5/algorithms/spectemplates/}) provided by SDSS, assisted with HSC images. Figure \ref{fig:bokeh} shows three examples. M stars are easy to classify since their spectra are dominated by TiO bands \citep{kirkpatrick1991standard}, making the spectra to be very rugged and easy to confirm (Figure \ref{fig:bokeh} top panel). However, stars such as K-type main sequence stars have very featureless spectrum, our classification is mainly based on the slope of the continuum and the color of HSC imaging (Figure \ref{fig:bokeh} middle panel, the slope matches, and the orange color agrees with our knowledge of K-type stars). Passive galaxies (or called early-type galaxy) are another challenging case due to the lack of emission lines. The identifiable features would be the Balmer breaks and other absorption lines. In the bottom panel of Figure \ref{fig:bokeh}, we show an example of a passive galaxy spectrum shifted to $z=0.328$. The positions of Mg absorption and 4000 \text{\AA} Balmer break match well with our data.

\par
In some cases, either due to the limitation of the wavelength coverage or S/N ratio, we were not able to obtain enough information to classify our candidates. The spectrum could either be featureless, hard to match to any known lines, or could just be heavily impacted by noise. The results of classification of these by-products are listed in Table \ref{tab:classification}.\par
\par
Regarding the classification scheme, there are two properties of the companions that we care about, with the first being AGN activity. If the object has emission lines, we use their width and line ratios to determine if the object is likely to be an AGN. Basically, when we find broad emission lines with width over 2000 $\mathrm{km\,s^{-1}}$ (measured via line fitting described in \ref{sec:Fitting}) in the spectrum, we will suggest this source to have AGN activity. For type2 sources, we utilize the BPT diagram. Another property worth noting is whether there are signs of an ongoing merger. This is mainly suggested from the HSC images, where disturbed features such as tidal tails are seen. For example, in Figure \ref{fig:2209}, and Figure \ref{fig:0219}. We found resolved candidates at same redshift that tend to have such features in their images, but when the redshift is higher than $\sim1$, the features are too faint to be observed. We note these two properties in column (5) and (6) in table \ref{tab:classification}, where "Y" for "yes", meaning having AGN activity or merger features, "N" for "no", meaning do not have and "P" for "possible", used in some ambiguous cases, e.g. when the data is located at the composite region of the BPT diagram, or the features are hard to tell from the images.\par

\begin{deluxetable*}{l|llccccl}
\tablenum{5}
\tablecaption{Spectral classification\label{tab:classification}}
\tablewidth{0pt}
\tablehead{
\colhead{} & \colhead{Name} & \colhead{Type} & \colhead{$z_A$} & \colhead{$z_B$} & \colhead{AGN activity} & \colhead{Merger} & \colhead{Features}\\
\colhead{} & \colhead{(1)} & \colhead{(2)} & \colhead{(3)} & \colhead{(4)} & \colhead{(5)} & \colhead{(6)} & \colhead{(7)}
}
\startdata
\hline
1 & 000439.97-000146.4 & Early-type galaxy & 0.5812 & 0.4037 & P & N & sodium absorption and H$\alpha$ emission\\
2 & 000508.37+010806.4 & K star & 1.3545 & 0 & N & N & spectrum shape and color\\
3 & 000837.66-010313.7 & Unclassified & 1.36 & \nodata & \nodata & \nodata & noisy\\
4 & 011935.29-002033.5 & M3 star & 0.8597 & 0 & N & N & spectrum shape and color\\
5 & 012716.17-003557.6 & Unclassified & 0.3611 & \nodata & N & P & visible disturbed feature in main source\\
6 & 020318.87-062321.3 & F star & 2.07 & 0 & N & N & spectrum shape and color\\
7 & 021352.67-021129.4 & type 1 Quasar & 2.778 & 2.844 & Y & N & possible broad C\,{\sc iv}\\
8 & 021930.51-055643.0 & Late-type galaxy & 0.2905 & 0.2905 & Y & Y & BPT diagram\\
9 & 022105.64-044101.5 & Galaxy & 0.1934 & 0.1949 & Y & Y & WHAN diagram\\
10 & 022159.71-014512.0 & F star & 2.343 & 0 & N & N & spectrum shape and color\\
11 & 084856.08+011540.0 & G star & 0.6457 & 0 & N & N & quasar+galaxy+star triple system\\
12 & 090347.33+002026.2 & M1 star & 0.4124 & 0 & N & N & absorption feature\\
13 & 094132.90+000731.1 & Galaxy & 0.4819 & 0.1355 & N & N & weak H$\alpha$\\
14 & 220228.65+004901.9 & M1 star & 1.445 & 0 & N & N & spectrum shape\\
15 & 220501.19+003122.8 & F star & 1.6364 & 0 & N & N & spectrum shape and absorption lines\\
16 & 220718.43-001723.1 & Galaxy & 0.7068 & 0.3875 & N & N & H$\alpha$ emission and sodium absorption\\
17 & 220811.56+023830.1 & F star & 3.172 & 0 & N & N & H$\alpha$ and Na absorption observed\\
18 & 221115.06-000030.9 & M1 star & 0.4769 & 0 & N & N & spectrum shape and color\\
19 & 221227.74+005140.5 & F star & 1.767 & 0 & N & N & spectrum shape and color\\
20 & 221759.58+060426.3 & Early-type galaxy & 2.569 & 0.328 & N & N & Mg absorption\\
21 & 222057.44+000329.8 & K type star & 2.265 & 0 & N & N & spectrum shape and color\\
22 & 222929.45+010438.4 & Early-type galaxy & 1.445 & 0.137 & N & N & sodium absorption\\
23 & 225147.82+001640.5 & type 1 Quasar & 0.4076 & 0.577 & Y & N & projected pair\\
24 & 230322.74-001438.2 & F star & 3.215 & 0 & N & N & spectrum shape and color\\
25 & 231152.90-001335.0 & Early-type galaxy & 0.3456 & 0.345 & N & P & possibly broad H$\alpha$\\
26 & 232853.40+011221.8 & Unclassified & 0.5265 & \nodata & \nodata & \nodata & extremely extended [O\,{\sc iii}] in main source\\
\hline
\enddata
\tablecomments{\\
Classification of target types via matching the spectroscopic data with SDSS templates.\\
Column (2): Types of the companions according to SDSS templates.\\
Column (3) \& (4): Redshifts of the two sources estimated via template matching.\\
Column (5): Whether the companion has AGN activity, Y: Yes, N: No, P: Potential.\\
Column (6): Whether they are a merger, based on HSC images.\\
Column (7): Observed evidences for classification.
}
\end{deluxetable*}

In the following sections, we provide details of these by-products that might have scientific interest beyond this work.

\subsection{Projected quasar pairs} \label{sec:Projected}
Differing from physically-associated pairs, there is no inevitability for projected quasar pairs to form, no more than by chance alignment. However, they can still provide useful information. For example, \cite{prochaska2013quasars} studied the H\,{\sc i} environment of the foreground quasars using the absorption features of background quasars based on a sample of 650 projected quasar pairs with proper separations between 30 kpc and 1 Mpc. Their follow-up study \cite{prochaska2014quasars} shows the similar method can also be appleid to study the C\,{\sc ii}$\lambda$1334 and C\,{\sc iv}$\lambda$1549 environment. Since the separations of our candidates are smaller than 30 kpc, these candidates would be helpful to study very central region of the quasars' environment, where an enhanced absorption is expected. In this work, we have discovered two new projected pair quasar systems at kpc-scale separation.

\subsubsection{SDSS J021352.67-021129.4}
\begin{figure*}[htp]
\centering
\includegraphics[width=1\textwidth]{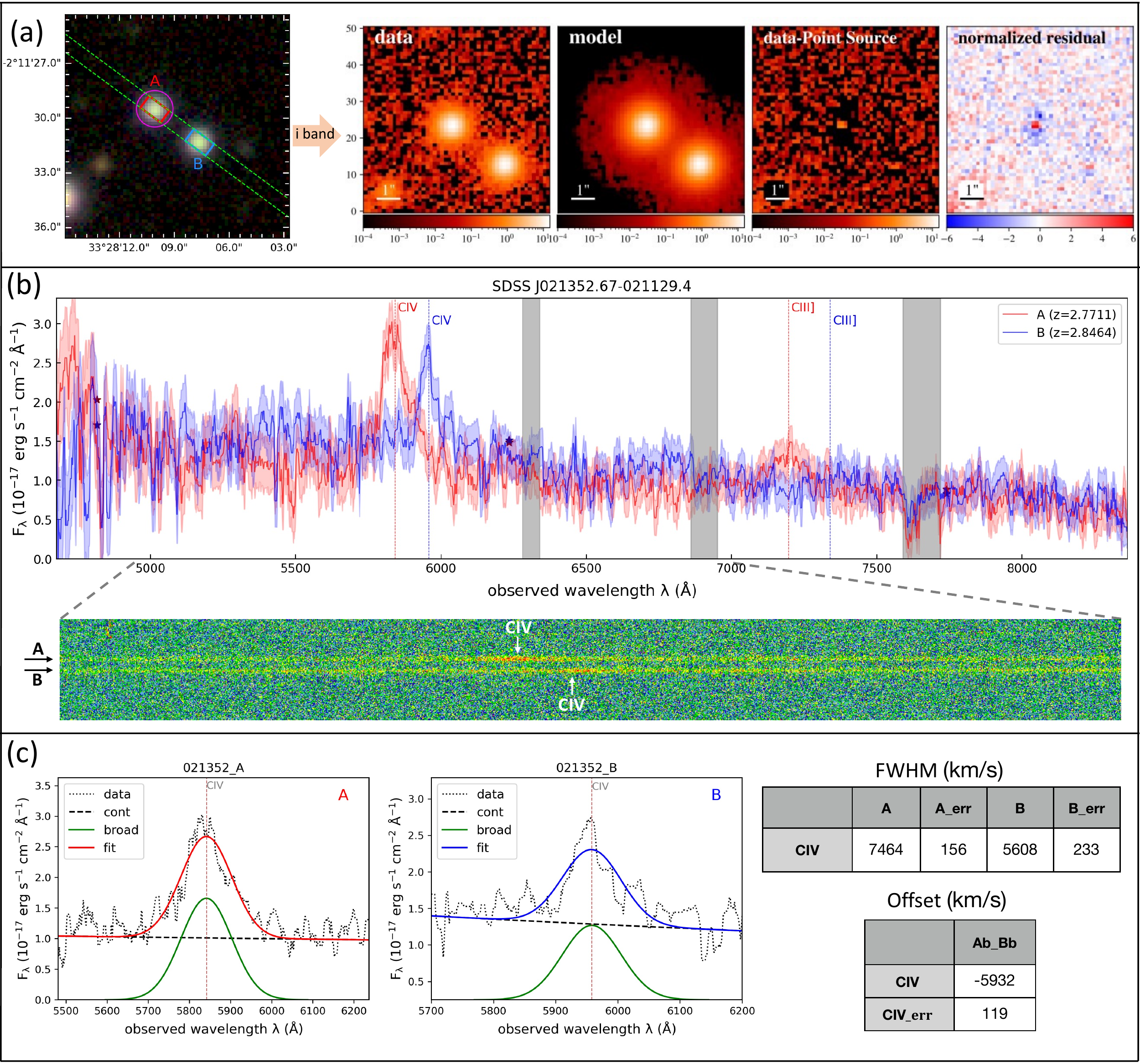}
\caption{A kaleidoscope of SDSS J0213-0211. Panel (a): HSC imaging overlaid with spectroscopic setups and decomposition results on $i$ band image. Panel (b): 1D and 2D spectrum, the 1D spectrum of each source is rescaled to match with the PSF magnitudes labeled with star symbols respectively. The grey shadows indicate the telluric absorption bands. Panel (c): Supplementary information. The upper row shows the line fitting result of source A, and the lower row shows source B. The tables show the FWHMs and offsets of the C\,{\sc iv} emission line.
\label{fig:021352}}
\end{figure*}
The SDSS BOSS program reports this target as a broad-line quasar at $z = 2.7794$. Both of the sources are very blue, below the stellar locus, and in the quasar region (Figure \ref{fig:SDSS_color}). The host galaxies are not resolved at this high redshift, thus two PSFs are enough to model their optical emission (Figure \ref{fig:021352}a).
\par
The FOCAS spectrum covers C\,{\sc iv} $\lambda$1549 and C\,{\sc iii}] $\lambda$1909. Unfortunately, the observing condition was not ideal with a seeing around $2^{\prime\prime}$ plus the target was very close to the moon. This resulted in a noisy profile and cut off at the blue end (Figure \ref{fig:021352}b). The redshift of source A is solid since Ly$\alpha$, C\,{\sc iv} $\lambda$1549 and C\,{\sc iii}] $\lambda$1909 are detected by SDSS and our own spectrum. However, the solution of source B is ambiguous. We are not very confident to say the strong emission line from the companion is C\,{\sc iv}, because there is no obvious C\,{\sc iii}] $\lambda$1909 feature at the corresponding position. We fit the line with a gaussian function (Figure \ref{fig:021352}c), that resulted in a velocity offset from component A of $5779 \pm 197\,\mathrm{km\ s^{-1}}$. If we consider the line to be C\,{\sc iv} $\lambda$1549, then it is shifted by $5932 \pm 119 \,\mathrm{km\ s^{-1}}$ relative to source A, hence we classify it as a projected quasar pair here. On the other hand, if the emission line is Mg\,{\sc ii} $\lambda$2798 or C\,{\sc iii}] $\lambda$1909, the redshit of source B would be 1.3021 and 2.3742 respectively. The C\,{\sc iii}] assumption is less likely to be true, because in that case, C\,{\sc iv} $\lambda$1549 is expected to appear at the blue end of the spectrum, which is actually absent.

\subsubsection{SDSS J225147.82+001640.5}
\begin{figure*}[htp]
\begin{centering}
\includegraphics[width=1\textwidth]{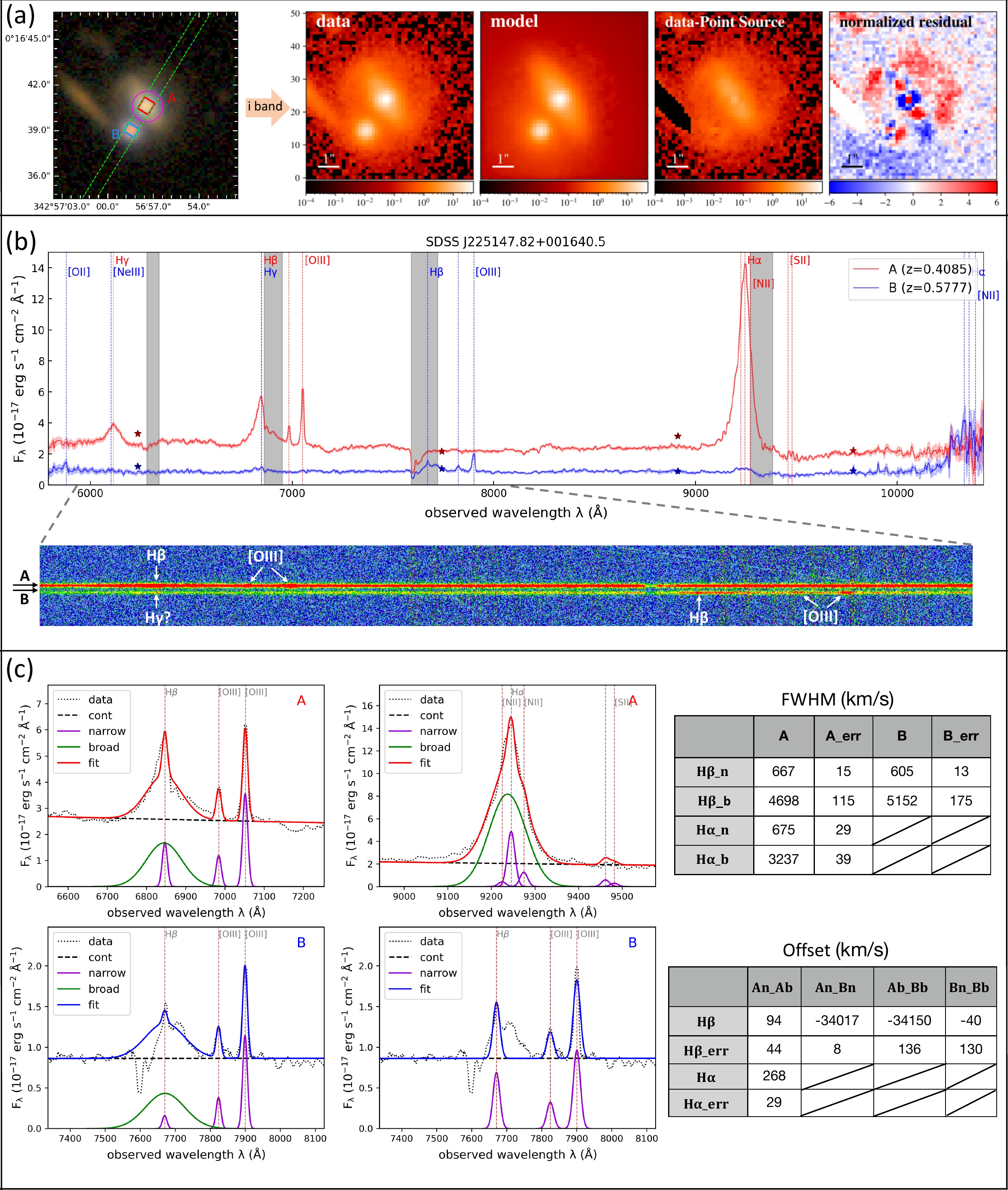}
\caption{A kaleidoscope of SDSS J2251+0016. Panel (a): HSC imaging overlaid with spectroscopic setups and decomposition results on $i$ band image. Panel (b): 1D and 2D spectrum, the 1D spectrum of each source is rescaled to match with the PSF magnitudes labeled with star symbols respectively. The grey shadows indicate the telluric absorption bands. Panel (c): Supplementary information. The upper row shows the line fitting result of source A, and the lower row shows source B. The tables show the FWHMs and offsets of the Balmer lines.
\label{fig:2251}}
\end{centering}
\end{figure*}
SDSS reports the northern source as a broad-line quasar at $z = 0.4096$. According to the HSC image, the main quasar is hosted by a face-on barred spiral galaxy, that is well resolved after we subract the point sources (Figure \ref{fig:2251}a data - point source).
\par
The FOCAS spectrum of this target covers the H$\alpha$ and H$\beta$ regions of the main quasar. However, we did not find similar features at similar positions in source B. Instead, they appear to be located at different wavelengths (Figure \ref{fig:2251}b). We find the H$\beta$ line is located near the 7600 \text{\AA} sky absorption feature, together with the [O\,{\sc iii}] doublet, which indicates a redshift of 0.5778. We tried to reconstruct the line profile that is affected by the sky absorption through line fitting in figure \ref{fig:2251}c. The bottom left figure shows a trial with a broad H$\beta$ component, that matches the redward part of the H$\beta$ profile well. Then we fit the profile with narrow components only, with the separation being fixed. We can see that with only narrow H$\beta$ the line profile is not fit adequately. This indicates the necessity of the broad component in this source (component B). The width of is $3408 \pm 203\,\mathrm{km\ s^{-1}}$. Therefore, we classify this target as a projected quasar pair. The H$\alpha$ line of the companion should be located at around 10250 \text{\AA}, which is likely buried in the noise at the red end of our spectrum. 

\subsection{Quasar-galaxy pairs} \label{sec:Galaxy}
In three cases, we found the companion to be a galaxy, without signs of a quasar, at the same redshift as the primary quasar. All of these cases are located at low redshift, and the HSC images are able to resolve the features of both galaxies. These systems show signs of interaction, indicating an on-going merging event.

\subsubsection{SDSS J021930.51-055643.0} \label{subsubsec:J0219}
\begin{figure*}[htp]
\centering
\includegraphics[width=1\textwidth]{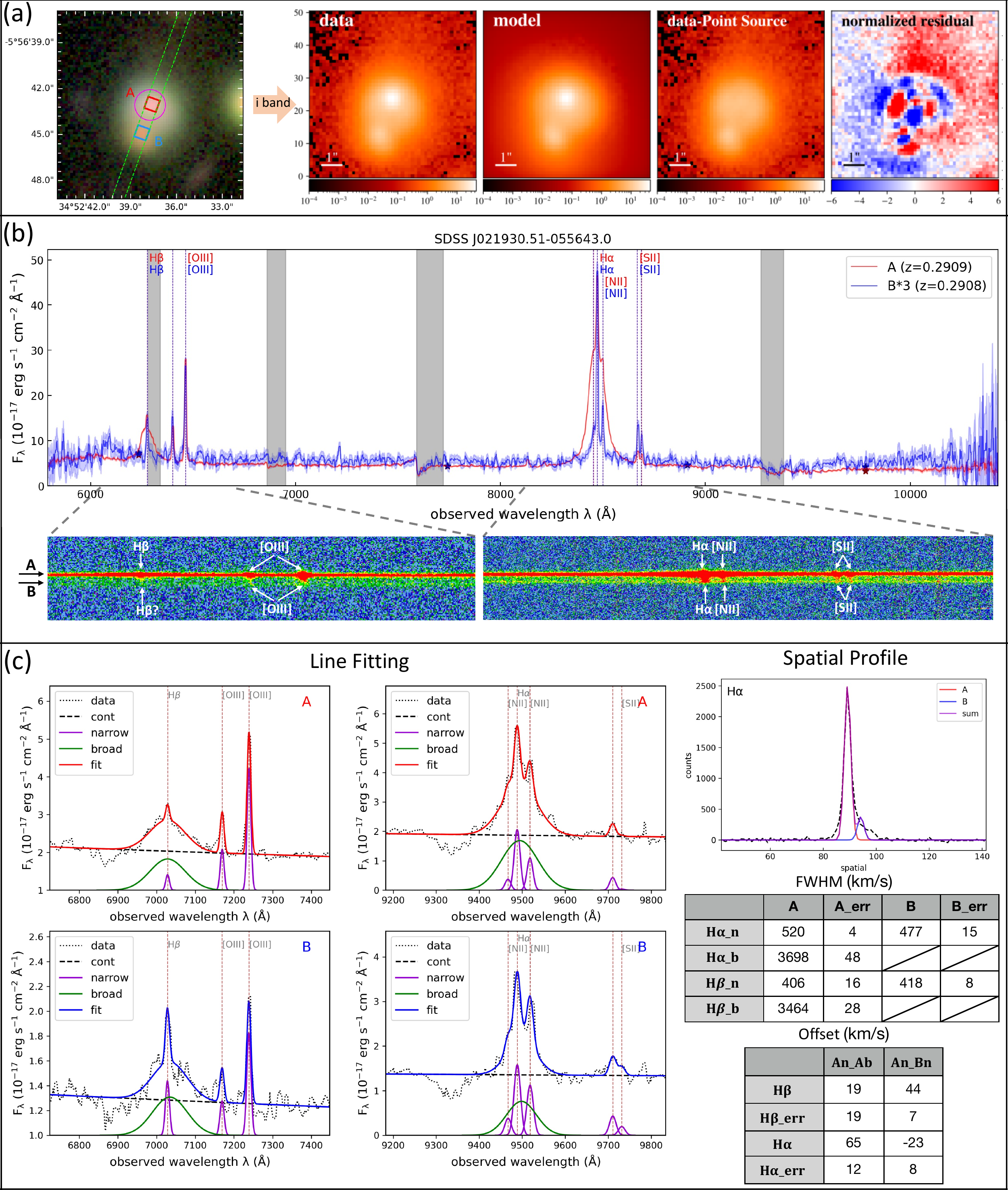}
\caption{A kaleidoscope of SDSS J0219-0556. Panel (a): HSC imaging overlaid with spectroscopic setups and decomposition results on $i$ band image. Panel (b): 1D and 2D spectrum, the 1D spectrum of each source is rescaled to match with the PSF magnitudes labeled with star symbols respectively. The grey shadows indicate the telluric absorption bands. Panel (c): Supplementary information. The upper row shows the line fitting result of source A, and the lower row shows source B. The upper right figure shows a shot of fitting to the spatial profile at the position of H$\alpha$. The tables show the FWHMs and offsets of the Balmer lines.
\label{fig:0219}}
\end{figure*}

\begin{figure*}[htp]
\begin{centering}
\includegraphics[width=0.95\textwidth]{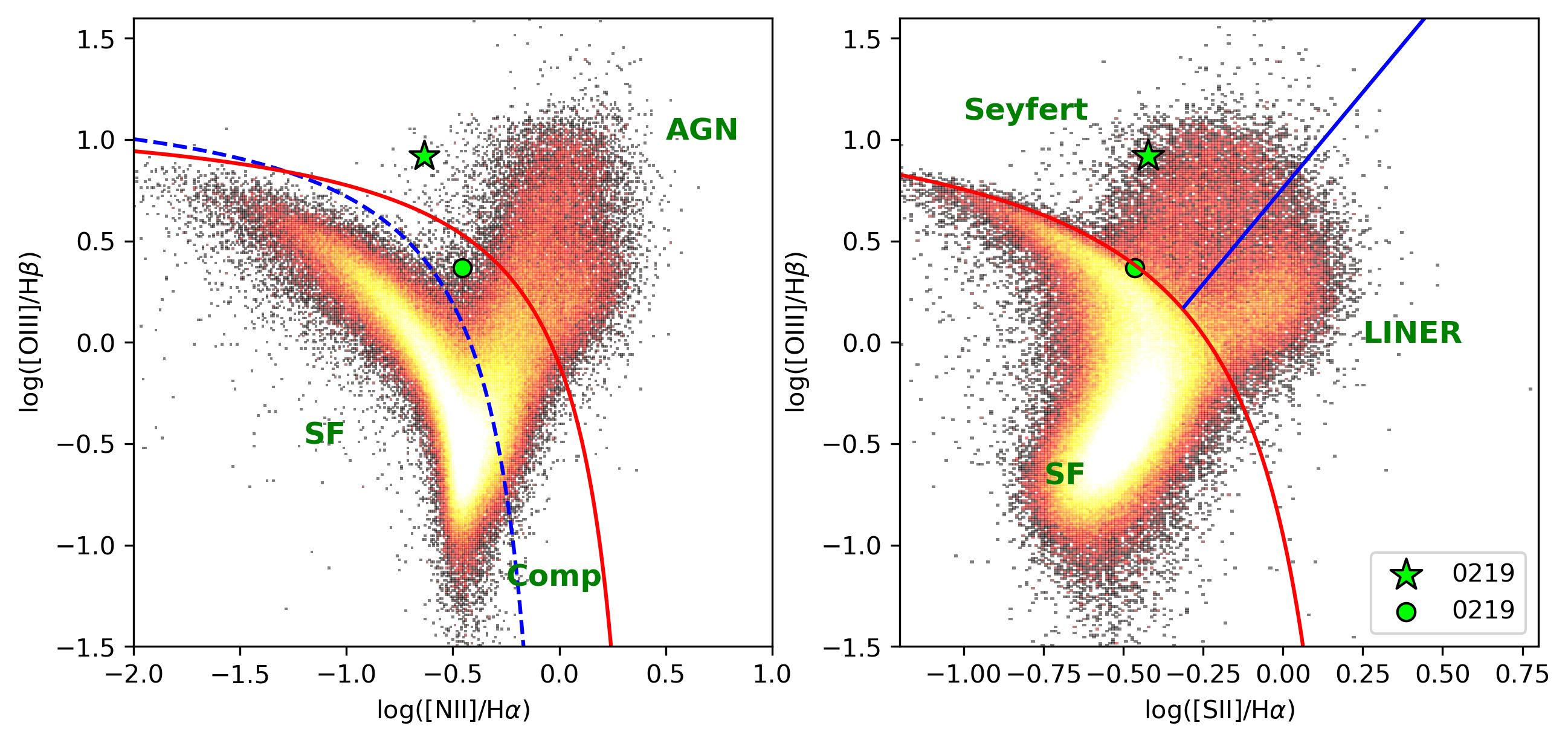}
\caption{BPT diagrams of SDSS J021930.51-055643.0. Upper panel for [O\,{\sc iii}]/H$\beta$ vs. [N\,{\sc ii}]/H$\alpha$, Lower panel for [O\,{\sc iii}]/H$\beta$ vs. [S\,{\sc ii}]/H$\alpha$. J0219 is plotted with an upper limit measurement of H$\alpha$. Background samples are 293,307 galaxies selected from SDSS DR16 \citep{ahumada202016th}.
\label{fig:0219_bpt}}
\end{centering}
\end{figure*}
SDSS classified this target as a broad-line quasar at $z = 0.2917$. The fiber is centered on the northern source as shown in left panel of Figure \ref{fig:0219}a. We decompose the emission based on model fitting using 2 PSFs and 2 Sersic functions (Figure \ref{fig:0219}a). The photometric magnitude of the companion is fainter by 1.2 mags on average in all five bands. We see that the host galaxies of these two sources are irregular given the strong residuals, thus indicating an ongoing interaction between them. 
\par

We provide a zoom-in view of the FOCAS spectrum of the H$\alpha$ and H$\beta$ regions in Figure \ref{fig:0219}b. We found that the H$\alpha$ emission of the main quasar is spatially extended and blends with emission from the companion. To disentangle the emission from each component, we perform an extraction of the profiles using \textbf{astropy} and \textbf{scipy} packages in python. We fit the spatial profile with two gaussian curves at each wavelength position. The aperture size is set to be a constant value for each source. Here we show the performance of this strategy at the position of H$\alpha$ as an example (Figure \ref{fig:0219}c Spatial Profile panel). The x-axis is the spatial axis along the vertical direction in the 2D spectrum in Figure \ref{fig:0219}b, and y-axis shows the counts at each pixel. Both sources are fairly well-described with a gaussian curve, we take the area under each curve as the photon counts at that wavelength position, then apply the flux calibration as usual. Besides H$\alpha$, we see other emission lines in the main quasar spectrum also tend to be more extended on the side of the companion than the other side. We suggest this to be an evidence of interaction between the two sources. 
\par
As a result of line fitting, we did not find significant broad components in both Balmer lines of the companion. We plot the line ratios of the pair in the BPT diagnostic diagram (Baldwin, Phillips \& Terlevich, \cite{baldwin1981classification, kewley2006host}, Figure \ref{fig:0219_bpt}), where the star marker is the main quasar, and circle is the companion. We have a very highly constrained measure of the lines from our spatial profile decomposition strategy, the error is smaller than the mark size. We find the companion is located at the composite region in both diagrams, thus hard to tell the origin of the emission. Here we classify it as a quasar-galaxy pair. Further confirmation is required with X-ray observations. At the distance of these two sources, the projected separation is 7.74 kpc at this redshift. 

\subsubsection{SDSS J022105.64-044101.5} \label{subsubsec:J0221}
\begin{figure*}[htp]
\begin{centering}
\includegraphics[width=1\textwidth]{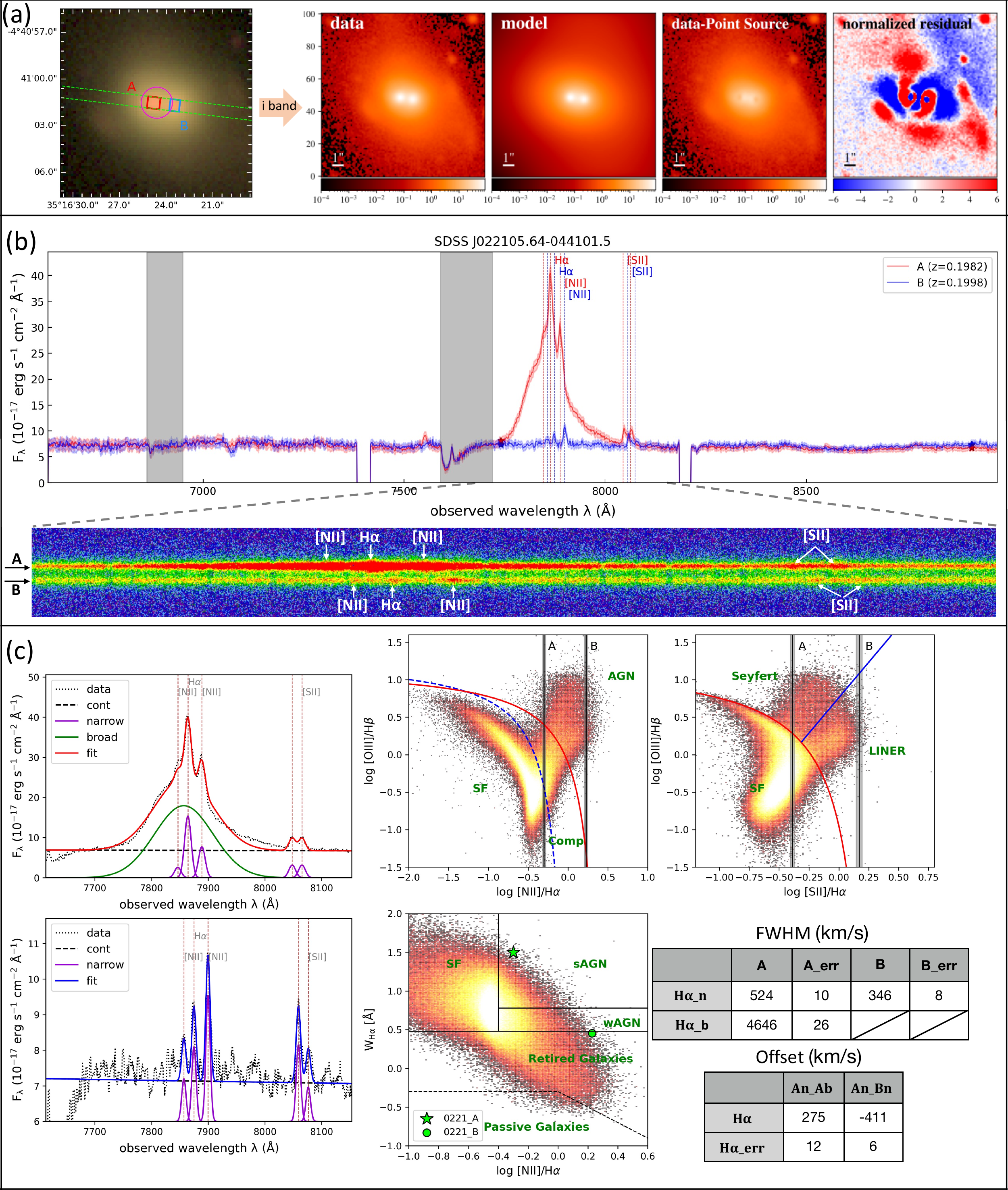}
\caption{A kaleidoscope of SDSS J0221-0441. Panel (a): HSC imaging overlaid with spectroscopic setups and decomposition results on $i$ band image. Panel (b): 1D and 2D spectrum, the 1D spectrum of each source is rescaled to match with the PSF magnitudes labeled with star symbols respectively. The grey shadows indicate the telluric absorption bands. Panel (c): Supplementary information. The upper row shows the line fitting result of source A, and the lower row shows source B. The diagnostics figures including BPT and WHAN help on the classification of the sources. The tables show the FWHMs and offsets of H$\alpha$.
\label{fig:0221}}
\end{centering}
\end{figure*}

SDSS reports the left source (Figure \ref{fig:0221}a) as a broad-line quasar at $z = 0.1986$. According to the HSC image, the sources have very similar color, located at the lower right side of the stellar locus (Figure \ref{fig:distance}), a relatively sparse region of quasars.
\par
H$\alpha$ is observed in our GMOS spectrum. The spectra of each component are well separated (Figure \ref{fig:0221}b). We see moderate H$\alpha$ emission in the spectrum of source B, with no broad component required in the fitting (Figure \ref{fig:0221}c left bottom panels). Different from J0219-0556, we do not have H$\beta$+[O\,{\sc iii}] information for J0221-0441, so we can only plot one axis on the BPT diagram (Figure \ref{fig:0221}c right upper panels). This object is located at the AGN and LINER region due to its strong [N\,{\sc ii}] emission. Therefore, our first conclusion was a LINER. As mentioned by \cite{cid2011comprehensive}, the LINER region of BPT diagrams consists of two distinct classes: weak active galactic nucleus (wAGNs) and retired galaxies (RGs). The latter are the result of ionization photons supplied by hot evolved low-mass stars (HOLMES) instead of AGN. One purpose of adding an extra WHAN diagram \citep{cid2011comprehensive} is to distinguish these two classes. Source B has a very high [N\,{\sc ii}]/H$\alpha$ ratio, up to 1.7, which places it at the right side of the diagram. The continuum level of this source is around 7.1$\times10^{-17}\,\mathrm{erg\ s^{-1}}\,\mathrm{cm^{-2}}\,\mathrm{\AA^{-1}}$, comparing to a flux of 20.3$\times10^{-17}\,\mathrm{erg\ s^{-1}}\,\mathrm{cm^{-2}}$ of H$\alpha$. This gives an equivalent width of 2.8 \text{\AA}, very close to the 3 \text{\AA} criterion suggested by \cite{cid2011comprehensive} to distinguish wAGN and RGs. Our data point almost falls on the borderline of these two classes (Figure \ref{fig:0221}c middle bottom figure). Of course, this diagnostic is not a dichotomy, but a continuous evolution from one to another. At the $W_{H\alpha}$ borderline, the AGN contributes between $\sim$ 1/3 and 2/3 of the ionizing power and decreases as $W_{H\alpha}$ decreases. In our case, this fraction is not negligible. We argue that there is an AGN contribution at some level to the lines, but according to the diagnostic diagram, we classify this as a galaxy.
\par
According to our model in Figure \ref{fig:0221}a, these two sources share one host galaxy, and they are very close to each other, separated by only $1^{\prime\prime}.23$, which projects to 4.04 kpc at redshift 0.199. Therefore, we suggest this to be a late-stage merging system with one of the SMBH not strongly activated yet or in a post\-AGN phase.\par

\subsubsection{SDSS J231152.90-001335.0} \label{subsubsec:J2311}
\begin{figure*}[htp]
\begin{centering}
\includegraphics[width=1\textwidth]{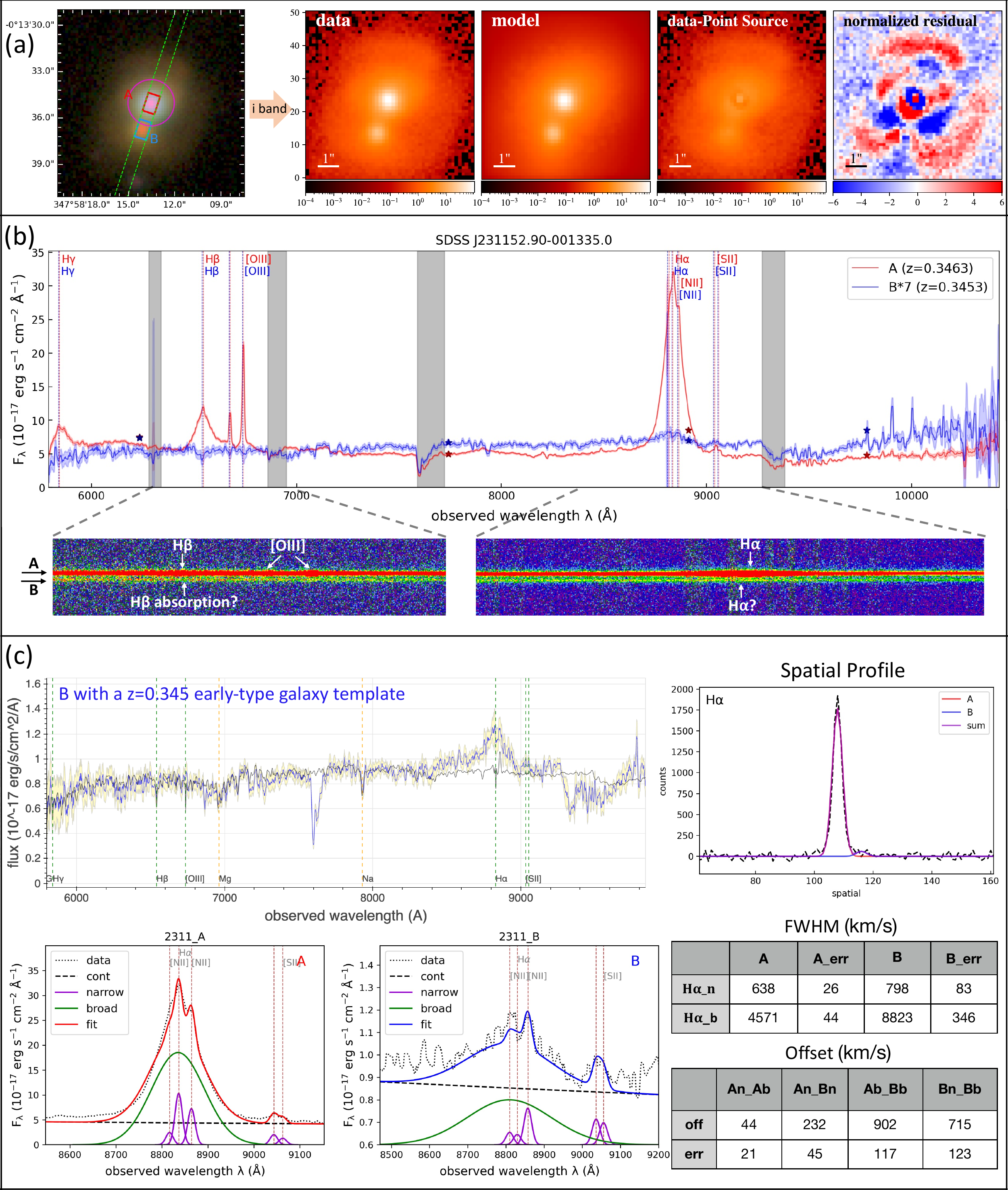}
\caption{A kaleidoscope of SDSS J2311-0013. Panel (a): image information of the HSC data and decomposition. Panel (b): 1D and 2D spectrum information. Panel (c): Supplementary information. The left upper figure shows a matching of a passive galaxy template (black curve) shifted to z=0.345 to the data of source B (blue curve). The line fitting result of both sources are plotted in the same row at bottom left. The right upper figure shows a fitting to the spatial profile at the position of H$\alpha$. The tables show the FWHMs and offsets of H$\alpha$.
\label{fig:2311}}
\end{centering}
\end{figure*}

SDSS reports the northern source (Figure \ref{fig:2311}a) to be a broad-line quasar at $z = 0.3473$. The fainter red companion resides to the southeast of the main source. The continuum level of the main source is about seven times that of the companion (Figure \ref{fig:2311}b).
\par
The redshift of the companion is supported by both absorption lines and a weak H$\alpha$ line. We matched the H$\beta$, Mg and Na absorption features in the spectrum to a passive galaxy template shifted to $z=0.345$, i.e. the same redshift as the main source (Figure \ref{fig:2311}c upper figure). At the corresponding position, we find a bump that is expected to be H$\alpha$ and fit it with a broad component (Figure \ref{fig:2311} panel (c) lower figures). While we matched the peak position of the two sources, the other components in the companion's spectrum are not convincing. It is hard to say whether a broad component is really required. If we believe there is a bump at H$\alpha$, then the velocity offset between the two sources is $418 \pm 23\,\mathrm{km\,s^{-1}}$. Our classification here is mainly based on the absorption lines, which leads to a result of a passive galaxy. 
\par
However, according to HSC imaging, the host galaxy of the main quasar overlaps with the companion source, so it is unavoidable that our spectrum of source B suffers from contamination. This issue cannot be well-resolved using the spatial profile fitting strategy (Figure \ref{fig:2311}c upper right panel) due to the low S/N ratio of source B. One result of this contamination is that, all the features, including the absorption lines and H$\alpha$ emission we see in the spectrum of source B, actually come from the host galaxy of the main quasar. In that case, our classification will no longer hold. Considering the morphology of the host galaxy of the main quasar, it is difficult for our decomposition model to reconstruct its features, as seen by the the normalized residual image. We suggest that it is likely an irregular galaxy. The separation of these two sources is $1^{\prime\prime}.75$, which projects to 8.6 kpc at redshift 0.346.

\end{document}